\documentclass[fleqn,usenatbib]{mnras}

\usepackage{newtxtext,newtxmath}

\usepackage[T1]{fontenc}
\usepackage{ae,aecompl}

\usepackage{graphicx}	
\usepackage{amsmath}	
\usepackage{amssymb}	

\title[Powerlaws from stochastic acceleration]{Powerlaw spectra from stochastic acceleration}

\author[M. Lemoine and M. Malkov]{
Martin Lemoine$^{1}$\thanks{e-mail: lemoine@iap.fr}
and Mikhail A. Malkov$^{2}$\thanks{e-mail: mmalkov@physics.ucsd.edu}
\\
$^{1}$Institut d'Astrophysique de Paris, CNRS -- Sorbonne Universit\'e, F-75014 Paris, France\\
$^{2}$University of California San Diego, La Jolla CA92093, USA\\
}

\pubyear{2020}

\begin{document}
\label{firstpage}
\pagerange{\pageref{firstpage}--\pageref{lastpage}}
\maketitle

\begin{abstract}
Numerical simulations of particle acceleration in magnetized turbulence have recently observed powerlaw spectra where pile-up distributions are rather expected. We interpret this as evidence for particle segregation based on acceleration rate, which is likely related to a non-trivial dependence of the efficacy of acceleration on phase space variables other than the momentum. We describe the corresponding transport in momentum space using continuous-time random walks, in which the time between two consecutive momentum jumps becomes a random variable. We show that powerlaws indeed emerge when the experimental (simulation) timescale does not encompass the full extent of the distribution of waiting times. We provide analytical solutions, which reproduce dedicated numerical Monte Carlo realizations of the stochastic process, as well as analytical approximations. Our results can be readily extrapolated for applications to astrophysical phenomenology.
\end{abstract}

\begin{keywords}
particle acceleration -- turbulence
\end{keywords}



\section{Introduction}\label{sec:introd}
In Fermi's original model of particle acceleration~\citep{49Fermi,1954ApJ...119....1F}, charged particles can increase their energy in a stochastic fashion through repeated interactions with moving magnetized structures. Magnetized turbulence has consequently been recognized as a natural source of suprathermal particles in space plasmas, and indeed, it may well be responsible for the non-thermal radiation emitted by a wide variety of astrophysical objects, from the Sun~\citep[{\it e.g.},][and references therein]{96Larosa,2004ApJ...610..550P,2004MNRAS.354..870S,2012ApJ...754..103B} to the remote Universe~\citep[{\it e.g.},][and references therein]{1977A&A....54....1L,1984A&A...136..227S,2000A&A...360..789S,2006A&A...453...47K,2007MNRAS.378..245B,2008ApJ...682..175P,2011ApJ...739...66T,18Asano}.

This broad interest has brought about a vast literature on the theoretical aspects of stochastic acceleration. Analytically, it has been addressed through calculations based on resonant wave-particle interactions~({\it e.g.}, \citealt{2002cra..book.....S} and references therein; for recent literature, see {\it e.g.},  \citealt{2006ApJ...638..811C}, \citealt{14Lynn},  \citealt{18Xu}, \citealt{Demidem19} and references therein), or on non-resonant interactions between particles and velocity structures~\citep[{\it e.g.},][]{1983ICRC....9..313B,1988SvAL...14..255P,1990A&A...236..519D,2003ApJ...595..195W,2004ApJ...603...23C,2010ApJ...713..475J,2017ApJ...846L..28X,2019PhRvD..99h3006L}. Stochastic acceleration has been probed by numerical experiments that simulate the transport of particles in synthetic turbulence~\citep{99Michalek,2007EL.....7840003P,09Sullivan,14Fatuzzo,Demidem19}, in full 3D MHD simulations \citep{2004ApJ...617..667D,2006ApJ...638..811C,2009ApJ...707..404L,2014ApJ...783..143D,14Lynn,2017PhRvL.119d5101I}, and more recently, in 3D kinetic (particle-in-cell or PIC) simulations \citep{2015PhRvL.114q5002W,17Zhdankin,18Zhdankin,2018MNRAS.474.2514Z,2018ApJ...867L..18Z,18Comisso,2019ApJ...886..122C,2020ApJ...893L...7W,2020ApJ...894..136T}. 

Astrophysical applications borrow the predicted or observed transport (diffusion) coefficients and rely on Fokker-Planck-type equations to determine the particle distribution function as a function of time, see {\it e.g.}, \citet{1984A&A...136..227S}, \citet{2006ApJ...647..539B}, \citet{2008ApJ...681.1725S}, \citet{2011JCAP...12..010M}. Pile-up distributions\footnote{that is,  quasi-Maxwellian distributions that concentrate most of the energy at the maximum momentum allowed by the age of the system.} then emerge as a generic feature of stochastic acceleration, in the absence of energy losses or particle escape. Yet, the aforementioned PIC simulations, which by construction work in a closed box and neglect energy losses, have produced distribution functions with extended soft powerlaw tails, in sharp contrast with those expectations. These results are of prime importance, because they cast into question a wealth of phenomenological applications to astrophysics. 

The generation of a powerlaw as the result of competition between energy gain and energy loss or escape can be regarded as the gist of Fermi-type acceleration. Hence, the simplest explanation for the emergence of a powerlaw in those simulations is the existence of some trapping mechanism that inhibits acceleration for a fraction of the particles -- at least, on the timescale of the simulations -- and thereby acts as an effective escape mechanism. In this paper, we develop this line of thought in order to interpret the results of those numerical experiments, having in mind their extrapolation to astrophysical cases of interest.

On a formal level, random walks with trapping times belongs to the class of continuous-time random walks~\citep{1965JMP.....6..167M}, see \citet{1990PhR...195..127B} for a general review, and \citet{1995PhRvE..51.4807B} for an application to anomalous transport in plasmas. These stochastic processes are characterized by a random, continuous time step, which is itself characterized by a probability density. If the distribution of jump time intervals has a finite mean $\langle\tau\rangle$, then diffusion is normal, meaning that the probability of undergoing $n$ jumps on timescale $t$ converges at large $n$ to the normal distribution, with $\langle n\rangle \simeq t/\langle\tau\rangle$. In the absence of energy losses or particle escape, the momentum distribution can be described by the Green function of a standard random walk with fixed time step $\langle\tau\rangle$ in the large-time limit. More generally speaking, the Fokker-Planck formalism can be used to describe the transport in momentum space. By contrast, heavy-tailed distributions, who do not possess a finite mean, characterize L\'evy flights and are more properly described by fractional transport equations, see for instance  \citet{2013ApJ...778...35Z}, \citet{2017A&A...607A...7Z}  and \citet{2017PhRvL.119d5101I}. 

In the following, we study these two classes of continuous-time random walks, and provide two toy models, which represent in our view the simplest models that can account for the emergence of a powerlaw on the finite timescale of numerical simulations. The first model assumes that the mean waiting time between two jumps in momenta is finite, but that the distribution of waiting times is such that for some particles, it allows acceleration on the simulation timescale, while for others, it does not. The second model assumes that the waiting time is distributed according to a one-sided stable (L\'evy) distribution with infinite mean waiting time.

We do not aim at elucidating the origin of this segregation here but suggest that it arises from a hidden dependency of the acceleration rate on phase space variables other than the particle momentum. Consider for instance the pitch-angle cosine $\mu$ of the particle, as defined with respect to the direction of the magnetic field line. It can be regarded as an internal degree of freedom that is averaged out when one treats the bulk of the suprathermal particle population, which one does implicitly when considering a Fokker-Planck equation in momentum space. If scattering is slow (on the simulation timescale) for some range of $\mu$, then particles in that range of $\mu$ effectively remain trapped in momentum space, given that scattering is a requisite of stochastic acceleration. As a possible realization, consider the interaction of particles with magnetic mirrors moving along the magnetic field lines: particles outside the loss cone bounce on the mirror and thus gain (or lose) energy, while particles inside the loss cone ignore the mirror and therefore undergo little energy gain/loss. This picture appears in qualitative agreement with the observation that high-energy particles are strongly peaked near $\mu=0$ in the simulations of \citet{2019ApJ...886..122C}, while low-energy particles rather show $\vert\mu\vert\sim1$. 

Alternatively, one may consider a situation in which the efficacy of acceleration is inhomogeneous in space, as suggested by some other simulations~\citep{2020ApJ...894..136T}. At each time step, a fraction of the particles happens to be in a region in which scattering, hence acceleration, is efficient, while the rest of particles mainly drift along the magnetic field lines. In this case, the internal degree of freedom that has been integrated out in deriving the Fokker-Planck equation is the position. Nonetheless, the general statistical description of the acceleration process as a continuous-time random walk remains legitimate.

Ultimately, one would like to study the full Fokker-Planck equation, including the dependence on the variables $\mu$ or $\boldsymbol{x}$, but this introduces by definition an infinite number of degrees of freedom describing the functional form of the transport coefficients. In this sense, our toy models provide the simplest approach to this problem.

Our paper is laid out as follows. We first discuss models with finite waiting time in Sec.~\ref{sec:mod1} and address the second case of L\'evy $\alpha-$stable distributions in Sec.~\ref{sec:mod2}. In Sec.~\ref{sec:disc}, we provide general comments and discuss how these models are modified when one accounts for escape losses. We provide conclusions in Sec.~\ref{sec:conc}. The diffusion coefficient in momentum space, $\left\langle \Delta p\Delta p\right\rangle/2\Delta t$ is written $D_{pp}$ and throughout, unless otherwise noted, it is assumed that $D_{pp}\propto p^2$, in accord with the results of the above PIC simulations, and with theoretical expectations.

\section{A binary model for stochastic acceleration}\label{sec:mod1}
\subsection{Analytical solution}\label{sec:mod1-1}
Continuous-time random walks possess the following formal solution, known as the Montroll-Weiss formula: if $\psi(\Delta t)$ denotes the distribution of waiting time $\Delta t$, and $\phi(\Delta \ln p)$ the distribution of (log) jump increments $\Delta \ln p$ in momentum space, then the probability density for observing a shift of log-momentum $\ln p$ at time $t$, $\mathcal P(\ln p,\,t)$, can be obtained from the inverse Fourier-Laplace transform
\begin{equation}
\mathcal P(\ln p,\,t)\,=\,\frac{1}{2\pi}\int_{-\infty}^{+\infty}{\rm d}\kappa\, 
e^{i \kappa \ln p} \frac{1}{2i\pi}\int_L{\rm d}\lambda\,
e^{\lambda t} \,\hat{\tilde{\mathcal P}}(\kappa;\,\lambda)\,,
\label{eq:FLtr}
\end{equation}
with~\citep{1965JMP.....6..167M}:
\begin{equation}
\hat{\tilde{\mathcal P}}(\kappa;\,\lambda)=\frac{1-\tilde\psi(\lambda)}{\lambda}\frac{1}{1-\tilde\psi(\lambda)\hat \phi(\kappa)}\,.
\label{eq:ctrwth}
\end{equation}
Here the hat symbol represents a Fourier transform from $\ln p$ to $\kappa$, and the tilde symbol a Laplace transform from $t$ to $\lambda$. In Eq.~(\ref{eq:FLtr}), $L$ stands for the Bromwich contour.

In practice, however, the calculations become prohibitive for non-trivial distribution functions, and one must rely on approximations. Consider for instance the following distributions, which characterize acceleration on two possible timescales $T_-$ and $T_+$ with (fixed) energy gain $g$:
\begin{align}
&\psi\left(\Delta t\right)\,=\, P_-\delta\left(\Delta t - T_-\right) + P_+\delta\left(\Delta t-T_+\right),\nonumber\\
&\phi\left(\Delta \ln p\right)\,=\, \delta\left(\Delta\ln p - g\right)\,,
\label{eq:dist1}
\end{align}
with $P_+=1-P_-$, corresponding to
\begin{align}
&\tilde\psi(\lambda)\,=\, P_- e^{-\lambda T_-} + P_+ e^{-\lambda T_+}\,,\nonumber\\
&\hat\phi(\kappa)\,=\, e^{-i \kappa g}\,.
\label{eq:dist2}
\end{align}
To simplify Eq.~(\ref{eq:ctrwth}), we take the limit $T_+\rightarrow +\infty$, which describes a situation in which $T_+$ is effectively much larger than the times $t$ on which we probe the distribution function, {\it e.g.}, the simulation timescale. Then $\hat{\tilde{\mathcal P}}(\kappa;\,\lambda)$ presents poles in $\lambda$ at $\lambda=0$ and $\lambda=T_-^{-1}\left(-i \kappa g  + \ln P_- + 2 i n \pi\right)$, for all $n\in \mathbb{Z}$. The former provides the late-time (stationary regime) scaling and we concentrate on it. We thus obtain the stationary distribution $\mathcal P_{\rm s}(\ln p)$ as
\begin{align}
\mathcal P_{\rm s}(\ln p)&\,=\,\frac{1}{2\pi}\int_{-\infty}^{+\infty}{\rm d}\kappa\, e^{i \kappa \ln p}\,\,\underset{\lambda=0}{\rm Res}\,\hat{\tilde{\mathcal P}}(\kappa;\,\lambda)\nonumber\\
&\,=\,\frac{1-P_-}{2\pi}\int_{-\infty}^{+\infty}{\rm d}\kappa\, \frac{e^{i \kappa \ln p}}{1-P_-e^{-i \kappa g}}\nonumber\\
&\,=\,\left(1-P_-\right)\sum_{n=0}^{+\infty}P_-^n\,\delta\left(\ln p - n g\right)\,\sim\,p^{\ln P_-/g}\,.
\label{eq:pstat}
\end{align}
Therefore, the spectrum $p^2 f(p)\propto \mathcal P_{\rm s}\,{\rm d}\ln p/{\rm d}p$ is a powerlaw with exponent $-1+\ln P_-/g$. We recover here the formula of \cite{1978MNRAS.182..147B} for particle acceleration in a process with energy gain $g$, and escape probability $1-P_-$, as indeed, the limit $T_+\rightarrow+\infty$  turns $1-P_-$ into a probability of escape from the acceleration process.

To generalize the above solution to a time-dependent regime, including a more realistic description of the diffusion in momentum, we resort to an alternative description of the problem, borrowing on previous work by \citet{2006ApJ...642..244M}. These authors studied the transport (in configuration and momentum space) of particles subject to interactions with nonlinear fronts (weak shocks) in the precursor of a strong shock. The population of particles was split into an ensemble of particles that were trapped and convected with the moving fronts, because of their small pitch-angle cosines, and another population of particles that could explore the train of moving fronts, because of their larger pitch-angle cosines. This general framework nicely applies to the situation at hand, and we thus break our distribution function into two populations at each time step: $f_0(p,\,t)$ characterizes the subset of particles for which acceleration is inhibited, while $f_1(p,\,t)$ represents the group that undergoes acceleration at some rate $\nu_{\rm acc}=D_{pp}/p^2\propto p^0$.

To complete the model, we also define the transition rate of population $1$ towards $0$ as $\nu_{10}$, and the transition probability from population 0 to population 1 as $\nu_{01}$. These frequencies characterize the rates at which particles can respectively scatter out of, or into, the acceleration region in phase space. We thus write the following transport equations in momentum space for both populations, as in \citet{2006ApJ...642..244M}, replacing however the regular energy gain with a diffusion operator:
\begin{align}
\frac{\partial}{\partial t} f_0(p,\,t)&\,=\,-\nu_{01}f_0(p,\,t) + \nu_{10}f_1(p,\,t),\nonumber\\
\frac{\partial}{\partial t} f_1(p,\,t)&\,=\,+\nu_{01}f_0(p,\,t) - \nu_{10}f_1(p,\,t)  \nonumber\\
&\quad\,+\,
\frac{1}{p^2}\frac{\partial}{\partial p}
\left\{D_{pp}\, p^2 \frac{\partial}{\partial p}f_1(p,\,t)\right\}\,.\nonumber\\
&
\label{eq:sysan}
\end{align}
From the point of view of acceleration, a trap is created if $\nu_{01}\ll\nu_{10}$ and the  timescales $t$ on which we probe the distribution satisfies $t\lesssim \nu_{01}^{-1}$. For PIC simulations, $t\sim\mathcal O(10L_{\rm max}/c)$ in terms of the maximum scale of the turbulence $L_{\rm max}$, hence the present interpretation suggests that $\nu_{01}L_{\rm max}/c\lesssim 0.1$.  The following assumes, for simplicity, but also in line with the scaling of $\nu_{\rm acc}$, that $\nu_{01}$ and $\nu_{10}$ are independent of $p$. 

We provide in App.~\ref{sec:appA} a full solution of the above system of equations in integral form. For the sake of commodity, we make it explicit here for generic initial conditions:
\begin{align}
f_0\left(p,\,t\right)&\,=\,e^{-\nu_{01}t}f_0(p,0)\nonumber\\
&\quad\,+\,\frac{1}{2\pi}\int_0^{+\infty}{\rm d}p_0\,\left(\frac{p}{p_0}\right)^{-3/2}\left\{\int_{-\lambda _+/\nu_{\rm acc}}^{\nu_{01}/\nu_{\rm acc}}+\int_{-\lambda _-/\nu_{\rm acc}}^{+\infty}\right\}{\rm d}s\,
\nonumber\\
&\quad\quad\quad\,\times\,e^{-s \nu_{\rm acc}t}\frac{\cos\left[\Sigma(s)\ln(p/p_0)\right]}
{\Sigma(s)}\,\frac{\Gamma_0\left(p_0;\,-s\nu_{\rm acc}\right)}{p_0}\,\nonumber\\
f_1\left(p,\,t\right)&\,=\,\frac{1}{2\pi}\int_0^{+\infty}{\rm d}p_0\,\left(\frac{p}{p_0}\right)^{-3/2}\left\{\int_{-\lambda _+/\nu_{\rm acc}}^{\nu_{01}/\nu_{\rm acc}}+\int_{-\lambda _-/\nu_{\rm acc}}^{+\infty}\right\}{\rm d}s\,
\nonumber\\
&\quad\quad\quad\,\times\,e^{-s \nu_{\rm acc}t}\frac{\cos\left[\Sigma(s)\ln(p/p_0)\right]}
{\Sigma(s)}\,\frac{\Gamma_1\left(p_0;\,-s\nu_{\rm acc}\right)}{p_0}\,,
\label{eq:ansolf}
\end{align}
where
\begin{align}
\Sigma(s)&\,=\,\sqrt{\frac{\left(s+\frac{\lambda_-}{\nu_{\rm acc}}\right)
\left(s+\frac{\lambda_+}{\nu_{\rm acc}}\right)}{s-\frac{\nu_{01}}{\nu_{\rm acc}}}}\,,\nonumber\\
\lambda _{\pm}&\,=\,\frac{\nu_{\rm acc}}{2}\Biggl\{-\left(\frac{\nu_{10}+\nu_{01}}{\nu_{\rm acc}}+\frac{9}{4}\right)
\nonumber\\&\quad\quad\quad
\pm\left[
\left(\frac{\nu_{10}-\nu_{01}}{\nu_{\rm acc}}+\frac{9}{4}\right)^2 + 4 \frac{\nu_{01}\nu_{10}}{\nu_{\rm acc}^2}\right]^{1/2}\Biggr\}\,,
\label{eq:defsigla}
\end{align}
and
\begin{align}
\Gamma_0(p_0;\,-s\nu_{\rm acc})&\,=\,\frac{\nu_{01}\nu_{10}}{\left(-s\nu_{\rm acc} +\nu_{01} \right)^2}f_0(p_0,0)  \nonumber\\
&\quad\,+\, \frac{\nu_{10}}{-s\nu_{\rm acc} +\nu_{01}}f_1(p_0,0)\nonumber\\
\Gamma_1(p_0;\,-s\nu_{\rm acc})&\,=\, \frac{\nu_{01}}{-s\nu_{\rm acc} +\nu_{01}}f_0(p_0,0)+f_1(p_0,0)\,,
\label{eq:defgam}
\end{align}
As expected, the distribution function is controlled by two parameters only, the ratios of $\nu_{01}$ and $\nu_{10}$ to $\nu_{{\rm acc}}$.

For $\nu_{01}\,\rightarrow\,0$, $f_0(p_0,0)\rightarrow 0$ and $f_1(p_0,0)\rightarrow f_1^0 p_0\delta\left(p-p_0\right)$, we obtain in particular,
\begin{align}
f_0\left(p,\,t\right)&\,=\,f_1^0\frac{\nu_{10}/\nu_{\rm acc}}{4\sqrt{\frac{\nu_{10}}{\nu_{\rm acc}}+\frac{9}{4}}}\,\left(\frac{p}{p_0}\right)^{-\frac{3}{2} - \epsilon_p\sqrt{\frac{9}{4}+\frac{\nu_{10}}{\nu_{\rm acc}}}}\nonumber\\
&\quad\times\left\{
\text{erfc}\left[\frac{\left\vert\log \left(p/p_0\right)\right\vert}{\sqrt{4\nu_{\rm acc} t}}-\sqrt{\left(\nu_{10}+\frac{9}{4}\nu_{\rm acc}\right)t}\right]\right.\nonumber\\
&\quad\quad-\left.\,\left(\frac{p}{p_0}\right)^{2\epsilon_p\sqrt{\frac{9}{4}+\frac{\nu_{10}}{\nu_{\rm acc}}}}\right. \nonumber\\
&\quad\quad\quad\quad\times\,\left.\text{erfc}\left[\frac{\left\vert\log \left(p/p_0\right)\right\vert}{\sqrt{4\nu_{\rm acc} t}}+\sqrt{\left(\nu_{10}+\frac{9}{4}\nu_{\rm acc}\right)t}\right]\right\}
\,.\nonumber\\
f_1\left(p,\,t\right)&\,=\,f_1^0\frac{1}{2\sqrt{\pi\nu_{\rm acc}t}}\left(\frac{p}{p_0}\right)^{-3/2}\,e^{-\left(\nu_{10} + \frac{9}{4}\nu_{\rm acc}\right)t - \frac{\ln(p/p_0)^2}{4\nu_{\rm acc}t}}\,.
\label{eq:f0f1s}
\end{align}
In these equations, $\epsilon_p=+1$ if $p>p_0$, and $\epsilon_p=-1$ if $p\leq p_0$.
The stationary distribution function can be obtained as $f_{\rm s}(p)\simeq\underset{t\rightarrow +\infty}{\rm lim}f_0(p,\,t)$. It can be read off the equation for $f_0(p,\,t)$ above, noting that the brackets containing the complementary error functions erfc tend to $2$ as $t\rightarrow+\infty$. This distribution function is thus characterized by a  powerlaw at high momenta $p>p_0$, with
\begin{equation}
p^2 f_{\rm s}(p)\,\propto\, p^{\frac{1}{2} - \sqrt{\frac{9}{4}+\frac{\nu_{10}}{\nu_{\rm acc}}}}\,.
\label{eq:pwls}
\end{equation}

\begin{figure}
\includegraphics[width=0.45\textwidth]{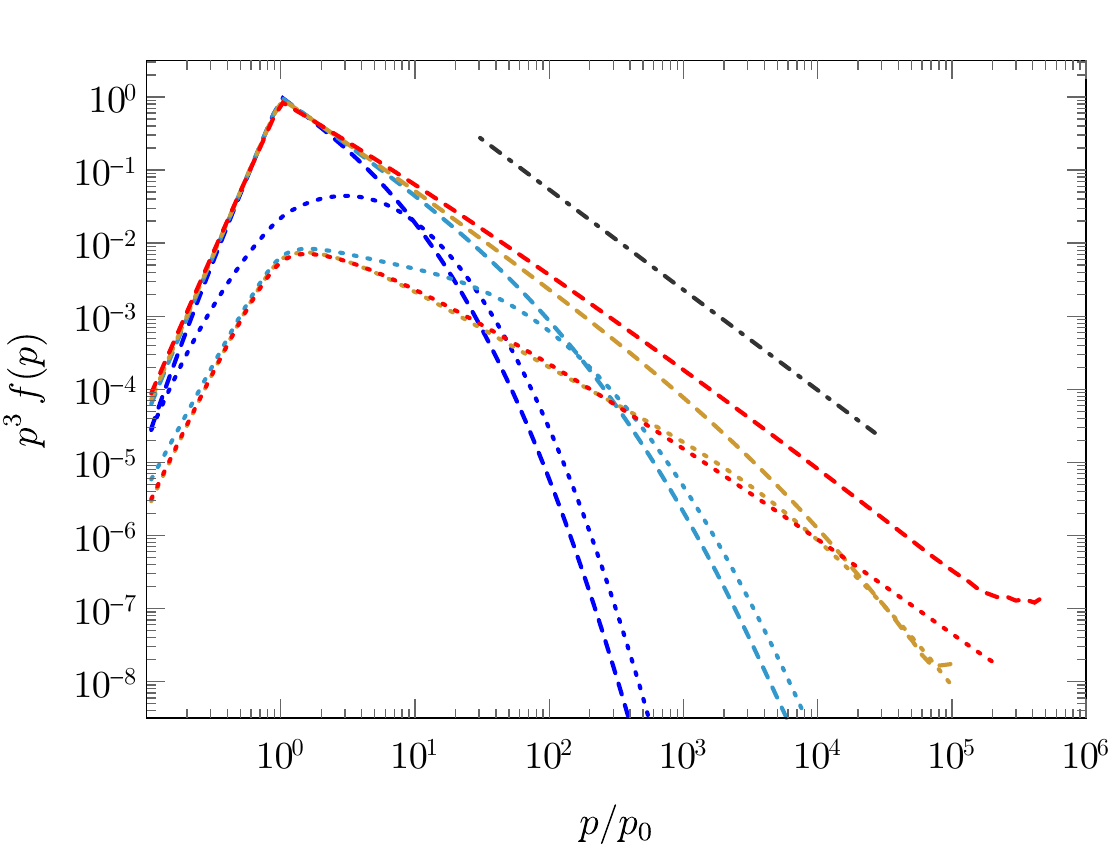}
 \caption{Distributions $p^3f(p,\,t)$ corresponding to the analytical solutions given in Eqs.~(\ref{eq:ansolf}) for the case in which $\nu_{01}/\nu_{\rm acc}=0.1$, $\nu_{10}/\nu_{\rm acc}=6$, with $f_0(p_0,0)=0$ and $f_1(p_0,0)=p_0\delta\left(p-p_0\right)$, plotted at different times $\nu_{\rm acc}t=0.4,\,0.8,\,1.5,\,3.$ (ordered from blue to red colors, or left to right). Dashed colored lines show the $f_0$ population, while dotted colored lines represent the $f_1$ population. On those timescales, $\nu_{01}t\,\ll\,1$, hence the solution takes the form of a powerlaw up to some momentum $p_{\rm max}$ that increases in time. This maximum momentum corresponds to the value of $p$ beyond which the spectrum turns over into a decaying exponential. For comparison, the dash-dotted gray line shows the powerlaw given by the analytical solution Eq.~(\ref{eq:pwls}), which assumes $\nu_{01}\rightarrow0$.   The plots for $\nu_{\rm acc}t=1.5,\,3.$ have been truncated at $p\gtrsim 10^5 p_0$ because of numerical errors in the evaluations of the integrals.
 \label{fig:ana1} }
\end{figure}

\begin{figure}
\includegraphics[width=0.45\textwidth]{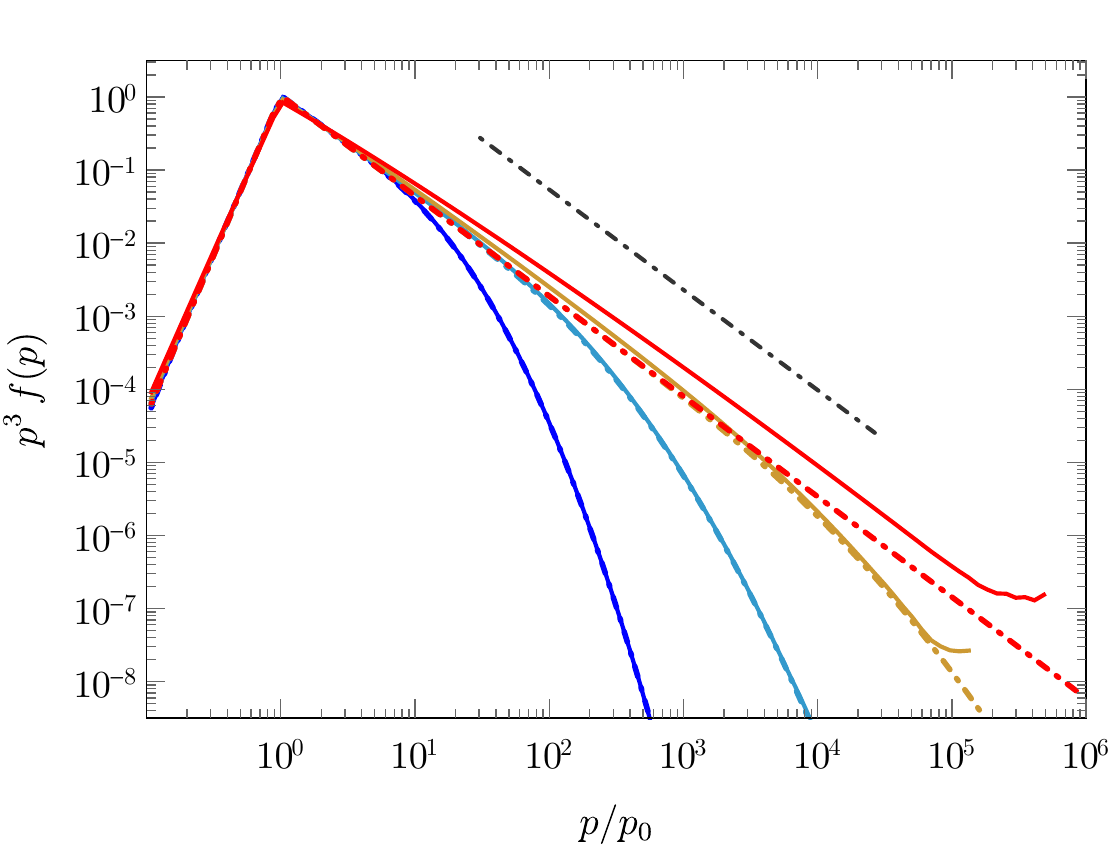}
 \caption{Same as Fig.~\ref{fig:ana1}, now comparing the sum $p^3\left[f_0(p,\,t)+f_1(p,\,t)\right]$ from the numerical evaluation of the full solution given by Eq.~(\ref{eq:ansolf}), represented with solid lines, with its approximate expression for $\nu_{01}\rightarrow 0$, as given by Eq.~(\ref{eq:f0f1s}), shown here in dash-dotted lines. For $\nu_{01}t\,\lesssim\,0.1$, the two are nearly identical. A difference of about a factor two can be seen at large momenta for $\nu_{01}t=0.3$.
 \label{fig:ana1b} }
\end{figure}

\begin{figure}
\includegraphics[width=0.45\textwidth]{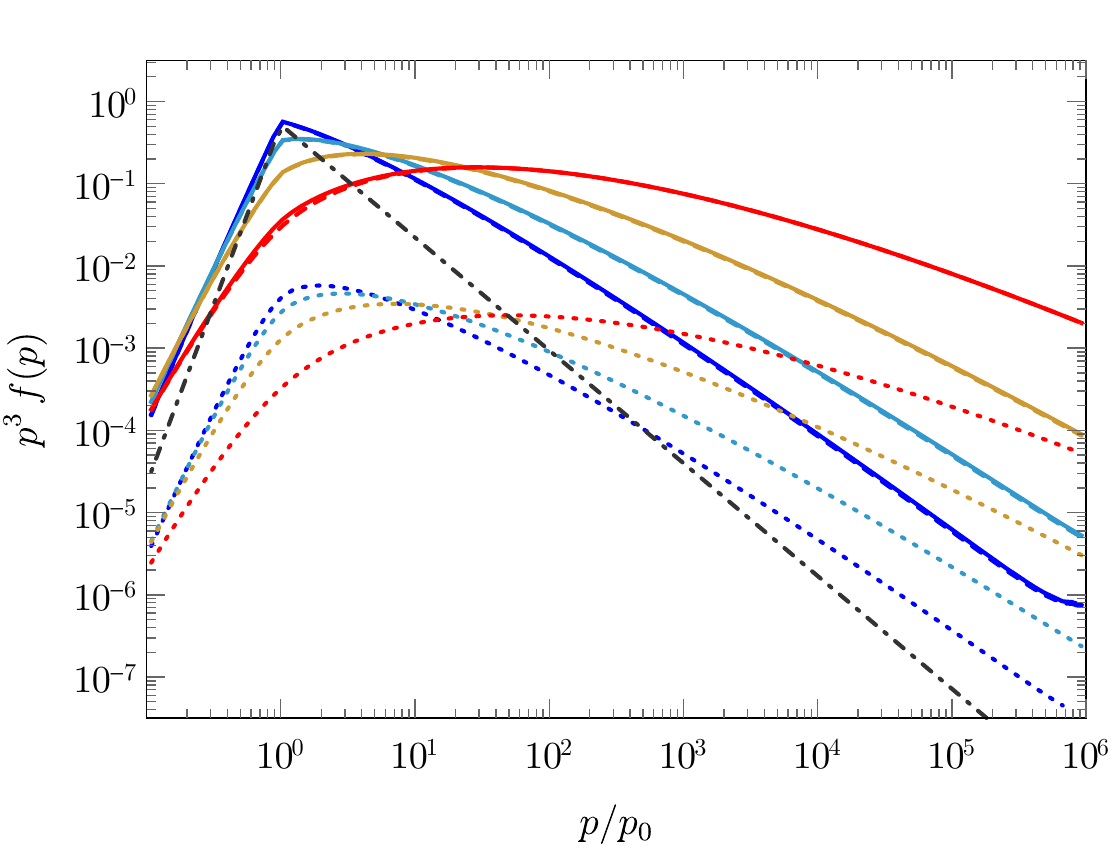}
 \caption{Same as Fig.~\ref{fig:ana1}, but now plotted at (large) times  $\nu_{\rm acc}t=10.,\,20.,\,40.,\,80.$, corresponding to $\nu_{01}t=1.,\,2.,,\,4.,\,8.$ (ordered from blue to red). In agreement with the theoretical expections, the solution now departs from a powerlaw to progressively turn into a pile-up distribution. Dashed colored lines show the $f_0$ population, while dotted colored lines represent the $f_1$ population. The gray dash-dotted line shows the analytical approximation for $\nu_{01}\rightarrow 0$, given in Eq.~(\ref{eq:f0f1s}); on these timescales, this solution has effectively reached its stationary state.
 \label{fig:ana2} }
\end{figure}

In Figs.~\ref{fig:ana1}, \ref{fig:ana1b} and \ref{fig:ana2}, we plot examples of such spectra, at different times, as measured in units of $\nu_{\rm acc}$. When $\nu_{01}t\,\ll\,1$, as is the case in Fig.~\ref{fig:ana1} and \ref{fig:ana1b}, the solution takes the form of a powerlaw at large momenta. As $\nu_{01}\,\ll\,\nu_{10}$ and $\nu_{01}\,\ll\,\nu_{\rm acc}$ in the present case, this index lies close to the theoretical value for $\nu_{01}\rightarrow0$, given in Eq.~(\ref{eq:pwls}). Specifically, Fig.~\ref{fig:ana1b} compares the full solution of Eq.~(\ref{eq:ansolf}) with its approximate expression for $\nu_{01}\rightarrow 0$ in Eq.~(\ref{eq:f0f1s}). It shows that the latter offers an excellent approximation to the former as long as $\nu_{01}t\,\lesssim\,0.1$.

As can be seen in Fig.~\ref{fig:ana1}, the full solution takes the form of an approximate powerlaw that turns over into a decreasing exponential beyond some momentum $p_{\rm max}$, whose scaling in time can be derived from the expression of $p^2f_1$ in Eq.~(\ref{eq:f0f1s}): the exponential part $\propto \exp\left[-\log(p/p_0)^2/(4\nu_{\rm acc} t)\right]$ doubles the powerlaw part $\propto (p/p_0)^{1/2}$ at 
$p_{\rm max}\sim p_0\exp\left(4\nu_{\rm acc}t\right)$. This value of $p_{\rm max}$  corresponds to the typical momentum that the particle would reach in a time $t$, in the absence of trapping.

\subsection{Random walk with trapping}\label{sec:mod1-2}
We now provide an alternative discretized description, which is more easily amenable to a numerical implementation, and which provides simple analytical estimates. Assume that, at each time step, of fixed extent $\Delta t$, a particle of the untrapped (accelerating) population can become trapped with probability $p_{10}$, but otherwise shifts in momentum, from $p$ to $p'=p+\Delta p$, according to the probability law 
\begin{equation} 
G_\nu(p'\vert p ,\Delta t)\,=\,\frac{\left(p'/p\right)^{1/2}}{p\sqrt{4\pi \nu\Delta t}}\, \exp\left[-\frac{9}{4}\nu\Delta t\,-\,\frac{\ln(p'/p)^2}{4\nu\Delta t}\right]\,, 
\label{eq:Green1} 
\end{equation}
with $\nu=\nu_{\rm acc}$. This corresponds to the propagator of the particle density $p^2f(p,\,t)$ for the Fokker-Planck equation, if $D_{pp}\propto p^2$, see Eq.~(\ref{eq:f0f1s}) in the limit $\nu_{10}\rightarrow 0$, or \citet{2006ApJ...647..539B}. 

Conversely, at each time step, a particle of the trapped (non-accelerating) population can become untrapped with probability $p_{01}$, without however gaining energy during $\Delta t$. The trapping probability is given by $p_{10}=1-\exp\left(-\nu_{10}\Delta t\right)\simeq\nu_{10}\Delta t$, while the untrapping probability (for trapped particles) is $p_{01}=1-\exp\left(-\nu_{01}\Delta t\right)\simeq\nu_{01}\Delta t$ for small $\Delta t$.

The distribution function at time $t$ can be then expressed as
\begin{equation}
p^2 f(p,\,t)\,=\, \sum_{n=1}^{+\infty}\,\frac{{\rm d}P_p}{{\rm d}p}\left(p,n\right)\,P_n\left(n;\,t\right)\,,
\label{eq:prob1}
\end{equation}
with ${\rm d}P_p/{\rm d}p$ the differential probability of reaching momentum $p$ after $n$ jumps of size $\Delta t$, and $P_n$ the probability of obtaining $n$ jumps in a time interval $t$. The former is simply $G_{\nu_{\rm acc}}(p\vert p_0,n\Delta t)$.

The probability of achieving at least $n$ jumps during $\Delta t$, {\it i.e.} $P_n(\geq n; t)$ corresponds to the probability that the total amount of time taken by these $n$ jumps does not exceed $t$, or equivalently, that the time the particle spends in traps does not exceed $t'=t - n \Delta t$, where $\Delta t$ is the length of a step notwithstanding trapping. If the particle encounters $m$ traps during these $n$ jumps, and if the waiting time for each jump is exponentially distributed with mean $\nu_{01}^{-1}$, the probability density ${\rm d}p_m/{\rm d}\tau$ of the sum $\tau$ of the $m$ waiting times reads
\begin{equation}
\frac{{\rm d}p_m}{{\rm d}\tau}\,=\,\frac{\tau^{m-1} \nu_{01}^m}{\Gamma(m)}e^{-\nu_{01}\tau}\,.
\end{equation}
The cumulative distribution up to time $t'\geq 0$ can thus be written as
\begin{equation}
\int_0^{t'}{\rm d}\tau\,\frac{{\rm d}p_m}{{\rm d}\tau}\,=\,
1-\frac{\Gamma\left[m,\nu_{01}t'\right]}{\Gamma(m)}\,.
\end{equation}
In the above equations, $\Gamma\left[a,b\right]$ stands for the incomplete Gamma function and $\Gamma(a)=\Gamma\left[a,0\right]$ for the Gamma function.
Consequently, the cumulative probability $P_n(\geq n;\, t)$ can be written
\begin{equation}
P_n(\geq n;\, t)\,=\,\sum_{m=0}^{n}C^n_m p_{10}^m\left(1-p_{10}\right)^{n-m}
\left\{1-\frac{\Gamma\left[m,\nu_{01}t'\right]}{\Gamma(m)}\right\}\Theta(t')\,,
\label{eq:Pn1}
\end{equation}
where $\Theta(t)$ represents the Heaviside function.
Finally, $P_n(n;\, t)=P_n(\geq n;\, t)-P_n(\geq n+1;\, t)$. This provides a formal solution to Eq.~(\ref{eq:prob1}).

To derive approximate expressions, we consider both  limits $\nu_{01}t'\gg1$ and $\nu_{01}t'\ll1$. The former describes a situation where particles can hop in and out of the traps frequently enough to homogeneize the acceleration process among the particle population, in which case we can expect to recover the standard propagator $G_{\nu'}(p\vert p_0,t-t_0)$ with some modified acceleration rate $\nu'$, to be determined. In the latter limit, however, a powerlaw should emerge and we seek to characterize its spectral index.

\subsubsection{Small traps, $\nu_{01}t'\gg1$}\label{sec:notrap}
For $\nu_{01}t'\gg1$, 
\begin{equation}
1-\frac{\Gamma\left[m,\nu_{01} t'\right]}{\Gamma(m)}\,\sim\,\Theta\left[\nu_{01} t' -m\right]\,.
\end{equation}
For large $n$, one can approximate the binomial distribution in Eq.~(\ref{eq:Pn1}) with a normal distribution:
\begin{align}
P_n(\geq n; t)&\,\simeq\,\int_0^n {\rm d}m \frac{\Theta\left[\nu_{01}t' -m\right]}{\left[2\pi n p_{10}\left(1-p_{10}\right)\right]^{1/2}}\exp\left[-\frac{\left(m-np_{10}\right)^2}{2np_{10}\left(1-p_{10}\right)}\right]\nonumber\\
&\,\simeq\,\Theta\left[\nu_{01}t'- np_{10}\right]\,,
\label{eq:Pn2}
\end{align}
the last equality following from the large $n$ limit, in which the gaussian for the variable $m/(np_{10})$ tends to a Dirac distribution centered on unity.

\begin{figure}
\includegraphics[width=0.45\textwidth]{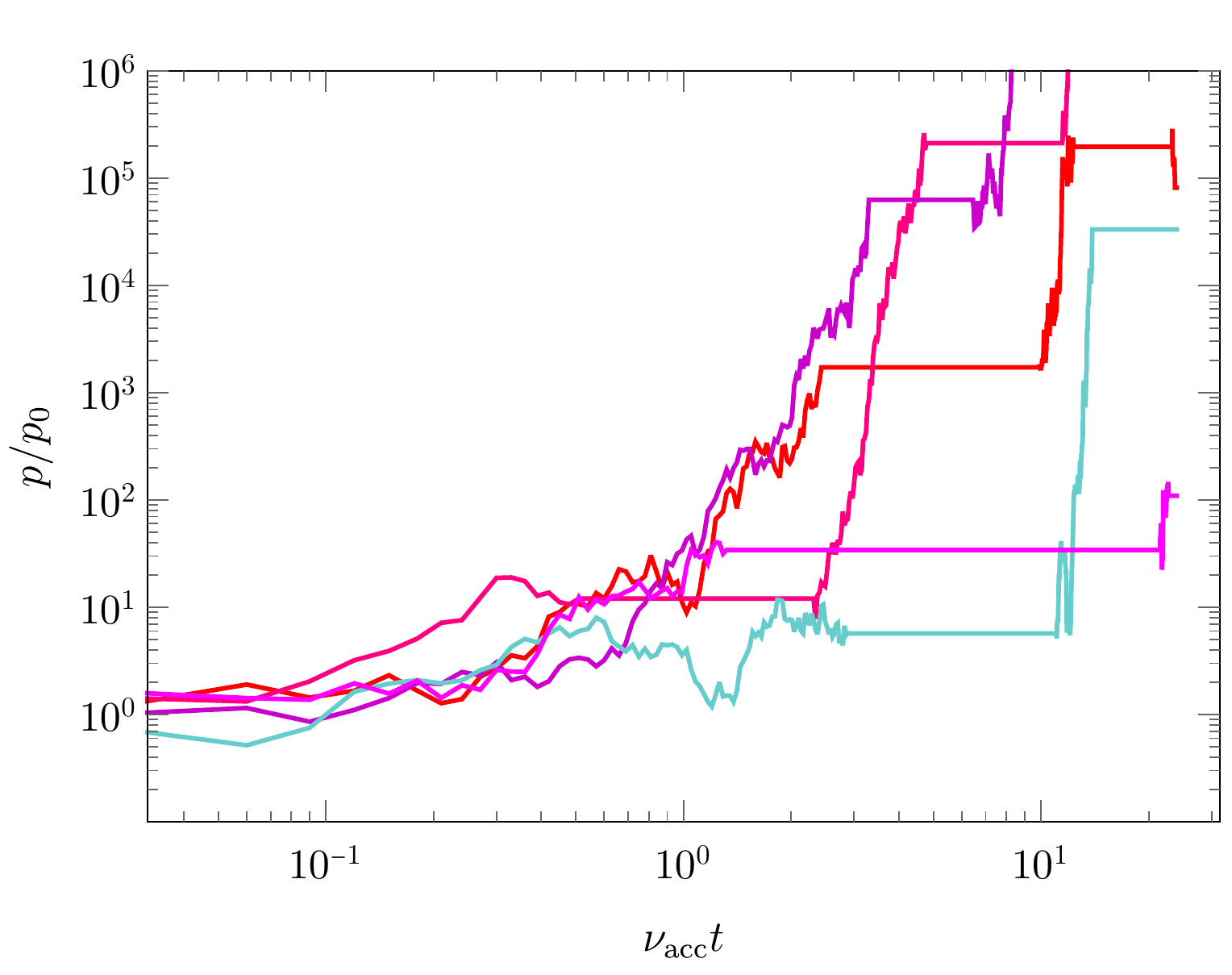}
 \caption{Typical trajectories in momentum space {\it vs} time, for 5 particles, with $\nu_{10}/\nu_{\rm acc}=0.5$ and $\nu_{01}/\nu_{\rm acc}=0.1$, as observed in a numerical Monte Carlo simulation of the discretized random walk of Sec.~\ref{sec:mod1-2}. Particles that get trapped at some time have effectively escaped the acceleration process on timescales $\lesssim \nu_{01}^{-1}$.
 \label{fig:num11} }
\end{figure}

Therefore,
\begin{equation}
P_n(\geq n;\,t)\,\simeq\,\Theta\left[\frac{t}{p_{10}\nu_{01}^{-1}+\Delta t}-n\right]\,,
\end{equation}
hence
\begin{equation}
P_n(n;\,t)\,\simeq\,\delta\left[n - \frac{t}{\langle\tau\rangle}\right]\,,
\end{equation}
where $\langle\tau\rangle=p_{10}\nu_{01}^{-1} + \Delta t$ gives the mean waiting time between two jumps, including the effect of trapping. The  distribution is then given by
\begin{align}
p^2f(p,\,t)&\,\simeq\,G_{\nu_{\rm acc}}\left(p\vert p_0,t\frac{\Delta t}{\langle\tau\rangle}\right)\nonumber\\
&\,=\,G_{\nu_{\rm acc}\Delta t/\langle \tau\rangle}\left(p\vert p_0,t\right)\,.
\label{eq:dNdp2}
\end{align}
Most of the particles are thus pushed to large momenta with $\langle p\rangle = p_0\exp(4\nu_{\rm acc}t\Delta t/\langle\tau\rangle)$, as in standard stochastic acceleration, except that the effective rate of acceleration has been reduced to $\nu_{\rm acc}\Delta t/\langle\tau\rangle \simeq \nu_{\rm acc}/\left(1 + \nu_{10}/\nu_{01}\right)$. Note that this recovery of the standard diffusion process is an illustration of the convergence theorem of continuous random walks with finite mean waiting time.

\begin{figure}
\includegraphics[width=0.45\textwidth]{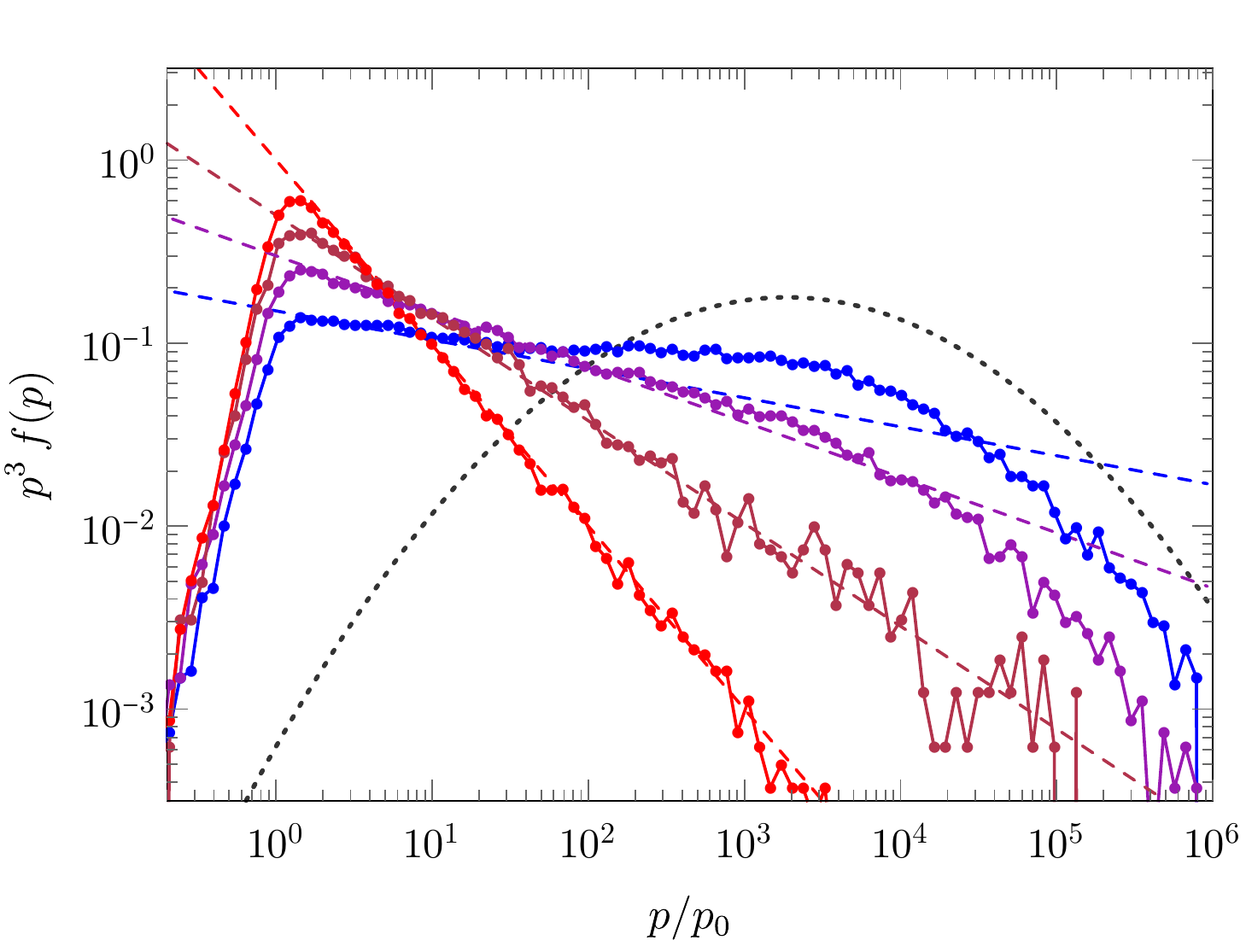}
 \caption{Distributions $p^3f(p,\,t)$ as a function of $p/p_0$ for $\nu_{10}/\nu_{\rm acc}=0.5,\,1,\,2,\,4$ (from hard to soft, or blue to red) and $\nu_{01}/\nu_{\rm acc}=0.01$. The spectra are plotted at time $t=2.5/\nu_{\rm acc}$. The dashed lines show the powerlaws expected from Eq.~(\ref{eq:dNdp3}), and the dotted line the expected spectrum in the absence of trapping, {\it i.e.} for $\nu_{10}/\nu_{\rm acc}\rightarrow0$.
 \label{fig:num12} }
\end{figure}

\begin{figure}
\includegraphics[width=0.45\textwidth]{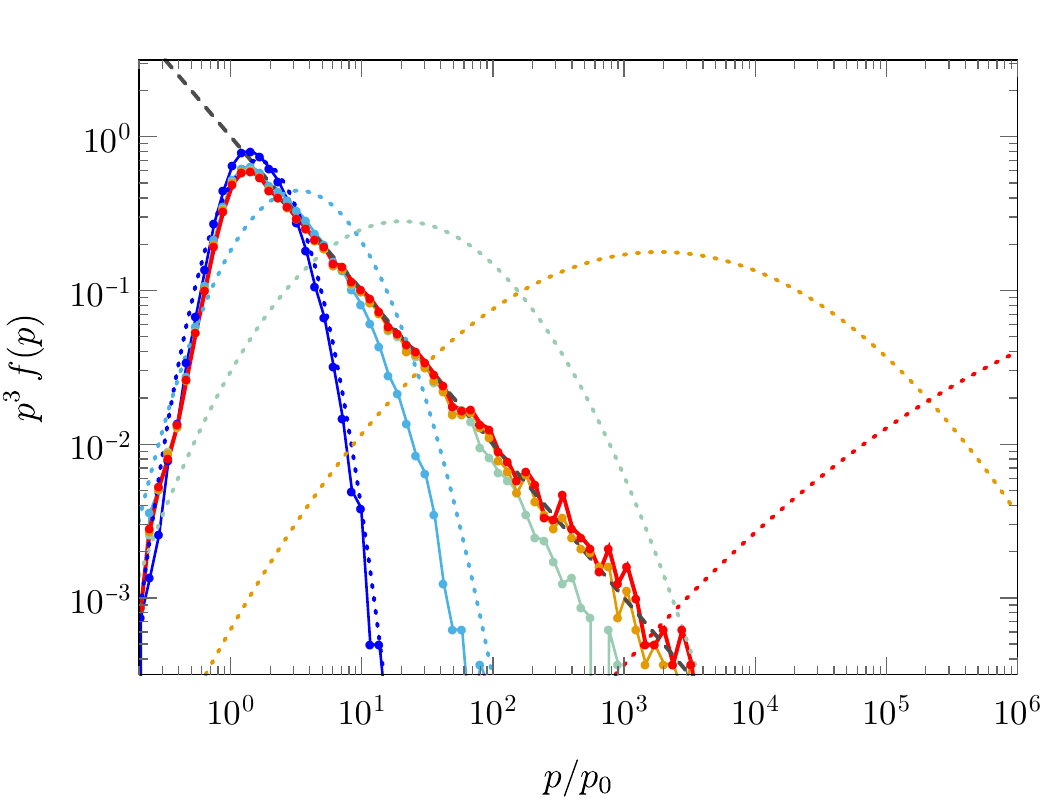}
 \caption{Spectra for $\nu_{10}/\nu_{\rm acc}=4$ and $\nu_{01}/\nu_{\rm acc}=0.01$, but plotted at different times: $\nu_{\rm acc}t\,=\,0.2,\,0.4,\,1.,\,2.5,\,6.3$ (ordered from blue to red, or left to right). Here $\nu_{01} t\ll1$ at all times, hence the spectra tend to an asymptotic powerlaw. The dotted lines show the expected spectra in the absence of trapping on the same timescales (from left to right). The dashed line shows the powerlaw $p^3 f\propto p^{-1}$ predicted by Eq.~(\ref{eq:dNdp3}).
 \label{fig:num13} }
\end{figure}

\begin{figure}
\includegraphics[width=0.45\textwidth]{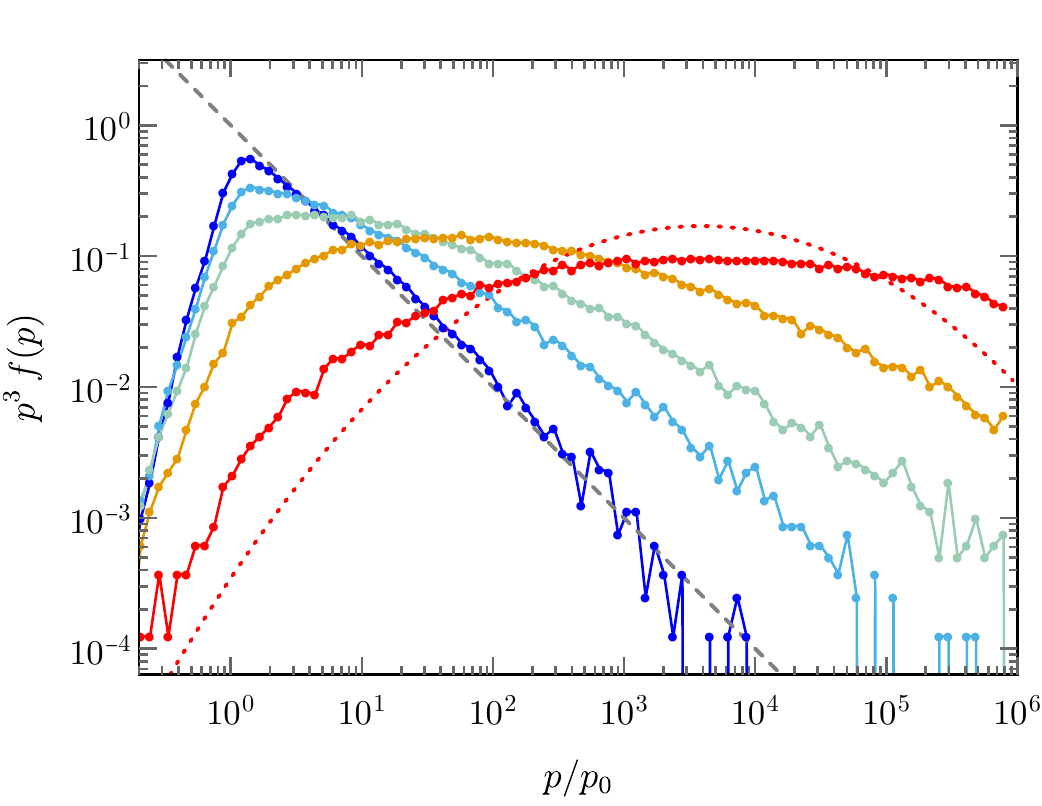}
 \caption{Same as Fig.~\ref{fig:num13}, but for $\nu_{10}/\nu_{\rm acc}=4$ and $\nu_{01}/\nu_{\rm acc}=0.2$, plotted at times $\nu_{01}t\,=\,0.3,\,0.6,\,1.2,\,2.6,\,5.6$ (ordered from left to right, or from blue to red). As $\nu_{01}t$ becomes larger than unity, the spectra depart from a powerlaw. The dotted line shows the analytical approximation given in Eq.~(\ref{eq:dNdp2}) for the last time step. The dashed line represents the powerlaw $p^3 f\propto p^{-1}$, which is expected at early times for this case.
 \label{fig:num14} }
\end{figure}

\subsubsection{Large traps, $\nu_{01}t'\ll1$}
In the opposite limit, $\nu_{01} t'\ll1$, the average number of traps encountered is small, because $\nu_{01}$ itself is small, meaning that the average trapping time far exceeds the timescale $t$. Then
\begin{equation}
1-\frac{\Gamma\left[m,\nu_{01}t'\right]}{\Gamma(m)}\,\simeq\,
\frac{\left(\nu_{01}t'\right)^m}{\Gamma(m+1)}\,.
\end{equation}
In this case, Eq.~(\ref{eq:Pn1}) is dominated by the terms at small $m$, giving
\begin{equation}
P_n(\geq n;\,\Delta t)\,\simeq\,\left(1-p_{10}\right)^n + n p_{10}\left(1-p_{10}\right)^{n-1}\nu_{01}t' + \ldots
\label{eq:pnlong}
\end{equation}
hence $P_n(n;\,\Delta t)\simeq-\ln\left(1-p_{10}\right)\left(1-p_{10}\right)^n$ to leading order in $\nu_{01}t'$. We are furthermore interested in the region of large momentum, which corrresponds to large $n$, hence we can approximate the sum in Eq.~(\ref{eq:prob1}) with an integral, giving
\begin{align}
p^2 f(p,\,t)&\,\simeq\,-\ln\left(1-p_{10}\right)\int_{1}^{+\infty}{\rm d}n\,e^{n\ln\left(1-p_{10}\right)}G_{\nu_{\rm acc}}\left(p\vert p_0;\,n\Delta t\right)\,,\nonumber\\
&\,=\,\left(\frac{p}{p_0}\right)^{\frac{1}{2}-\sqrt{\frac{9}{4}+\frac{\nu_{10}}{\nu_{\rm acc}}}}\,\left[\ldots\right]\,,
\label{eq:dNdp3}
\end{align}
noting that $\ln\left(1-p_{10}\right)=-\nu_{10}\Delta t$. The terms in brackets that have not been explicited match exactly the remainder of the solution for $f_0(p,\,t)$, as written in Eq.~(\ref{eq:f0f1s}). We thus obtain the same powerlaw scaling as before, to leading order in $\nu_{01}t'$.

In Figs.~\ref{fig:num11}, \ref{fig:num12}, \ref{fig:num13} and \ref{fig:num14}, we provide illustrations of this discretized process, as obtained from numerical Monte Carlo calculations, for various choices of parameters. Figure~\ref{fig:num11} shows the typical trajectories of particles in the presence of traps and illustrates how those particles that escape the traps can reach very high energies, while those that get trapped at some given time effectively escape the acceleration process, at least on timescale $\sim \nu_{01}^{-1}$. Note that the exponential scaling of $p$ as a function of $t$  -- which is preserved on a log-log plot -- is typical of the dependence $D_{pp}\propto p^2$, which we have assumed here.

Figure~\ref{fig:num12} plots spectra at a given time for different values of $\nu_{10}/\nu_{\rm acc}$ and compares them to the analytical approximation Eq.~(\ref{eq:dNdp3}), which applies in this case since $\nu_{01}t\,\ll\,1$. The correspondence is quite satisfactory. Figure~\ref{fig:num13} shows spectra at different times, for given values of $\nu_{10}/\nu_{\rm acc}$ and $\nu_{01}\nu_{\rm acc}$, with $\nu_{01}t\ll1$. This figure confirms that the spectrum takes the form of a powerlaw, whose spectral index does not change in time on those timescales. Finally, Fig.~\ref{fig:num14} shows how such spectra evolve on longer timescales, in particular $\nu_{01}t\gtrsim1$, slowly departing from a powerlaw to converge towards the (pile-up form) solution without traps, given in Eq.~(\ref{eq:dNdp2}).

\section{Stochastic acceleration with L\'evy traps}\label{sec:mod2}
L\'evy flights are characterized by heavy-tailed distributions whose mean (or r.m.s.) waiting time is infinite. Among this class of distributions, the family of one-sided stable distribution functions behaves as an attractor for the sums of (identically distributed) random variables, in analogy with the normal distribution for finite variance. For a positive real variable (time), these distributions $L_\alpha(\hat t)$ are defined by their characteristic function $\tilde L_\alpha(\hat\omega)$, which is itself characterized by a real parameter $\alpha\in ]0,1[$, 
\begin{align}
\tilde L_\alpha(\hat\omega)&\,\equiv\,\int {\rm d}\hat t\,e^{i\hat \omega \hat t}\,L_\alpha(\hat t)
\,=\,\exp\left[- i \varsigma\vert \hat \omega\vert^\alpha\right]\,,
\label{eq:Levy1}
\end{align}
with $\varsigma\,=\,\exp\left(-i \alpha\pi/2\right)$, and $\hat t=t/\Delta t$; $\Delta t$ represents a reference timescale for the timestep duration. In the large time limit, $L_\alpha(\hat t)\propto \hat t^{-1-\alpha}$, hence the average waiting time is indeed infinite if $\alpha <1$. In the following, we restrict our study to this family of distribution functions.

\begin{figure}
\includegraphics[width=0.45\textwidth]{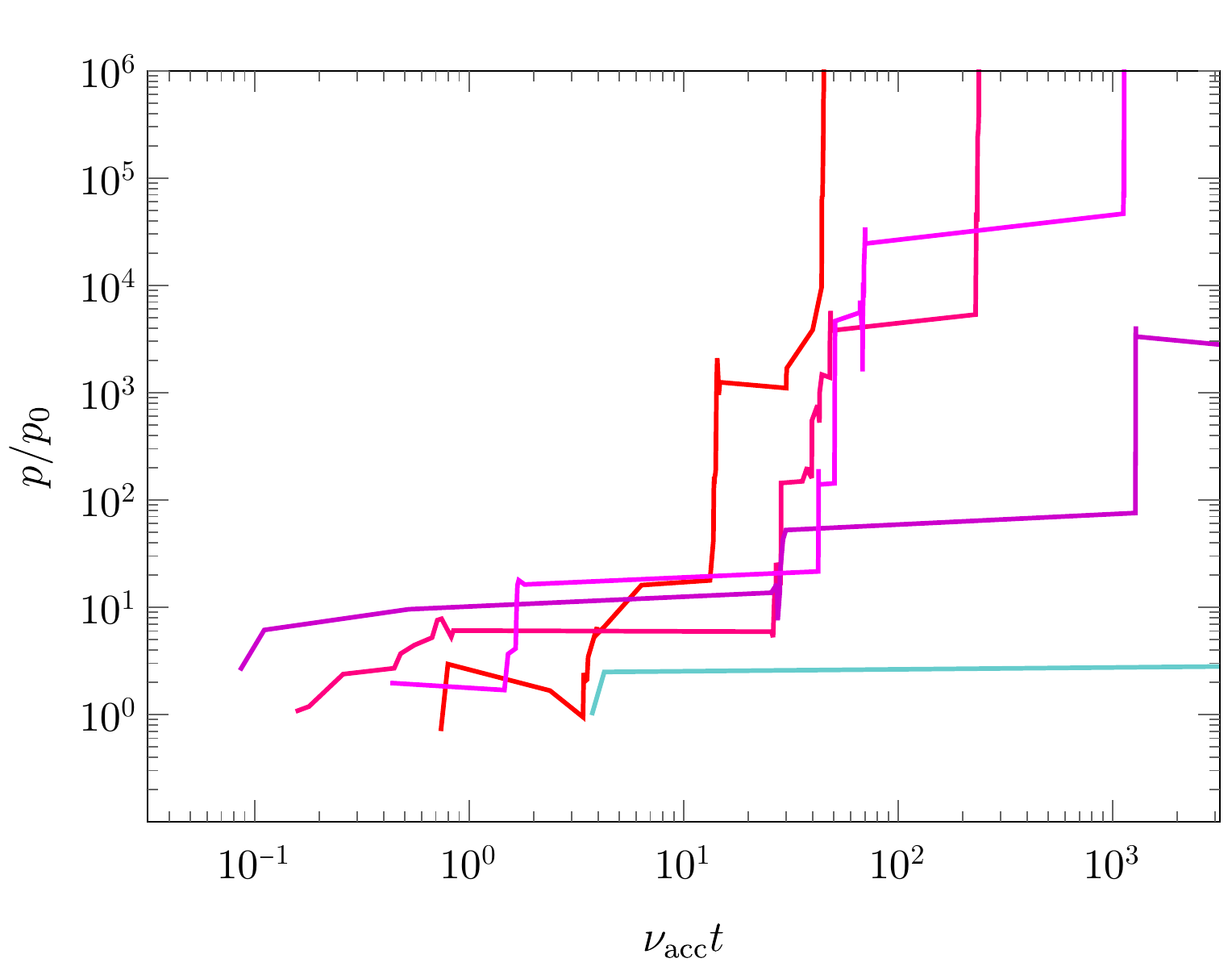}
 \caption{Typical trajectories in momentum space {\it vs} time, for 5 particles, for a Monte Carlo realization of L\'evy flights characterized by a distribution $L_\alpha(\hat t)$ with $\alpha=0.5$ and $\Delta t\nu_{\rm acc}=0.1$. 
 \label{fig:num21} }
\end{figure}

Consider therefore the following setup: at each step $n$, the momentum of the particle jumps by a random quantity $\Delta p$, which is distributed according to Eq.~(\ref{eq:Green1}) as before. However, before entering a new cycle of energy gain, the particle lingers over a time interval $\hat t \Delta t$, with $\hat t$ distributed according to $L_\alpha(\hat t)$. We should stress that Eq.~(\ref{eq:Green1}) computes $\Delta p$ assuming that the particle undergoes acceleration over a time interval $\Delta t$, while the actual time spent between two cycles, in the present description, is $\hat t\Delta t$, and $\hat t$ can in principle be smaller than unity. However, $L_\alpha(\hat t)$ goes rapidly to zero for values $\hat t$ smaller than unity, so that this choice does not impact our conclusions. Some example trajectories in momentum space are shown in Fig.~\ref{fig:num21} for $\alpha=0.5$ and $\nu_{\rm acc}\Delta t=0.1$.

\begin{figure}
\includegraphics[width=0.45\textwidth]{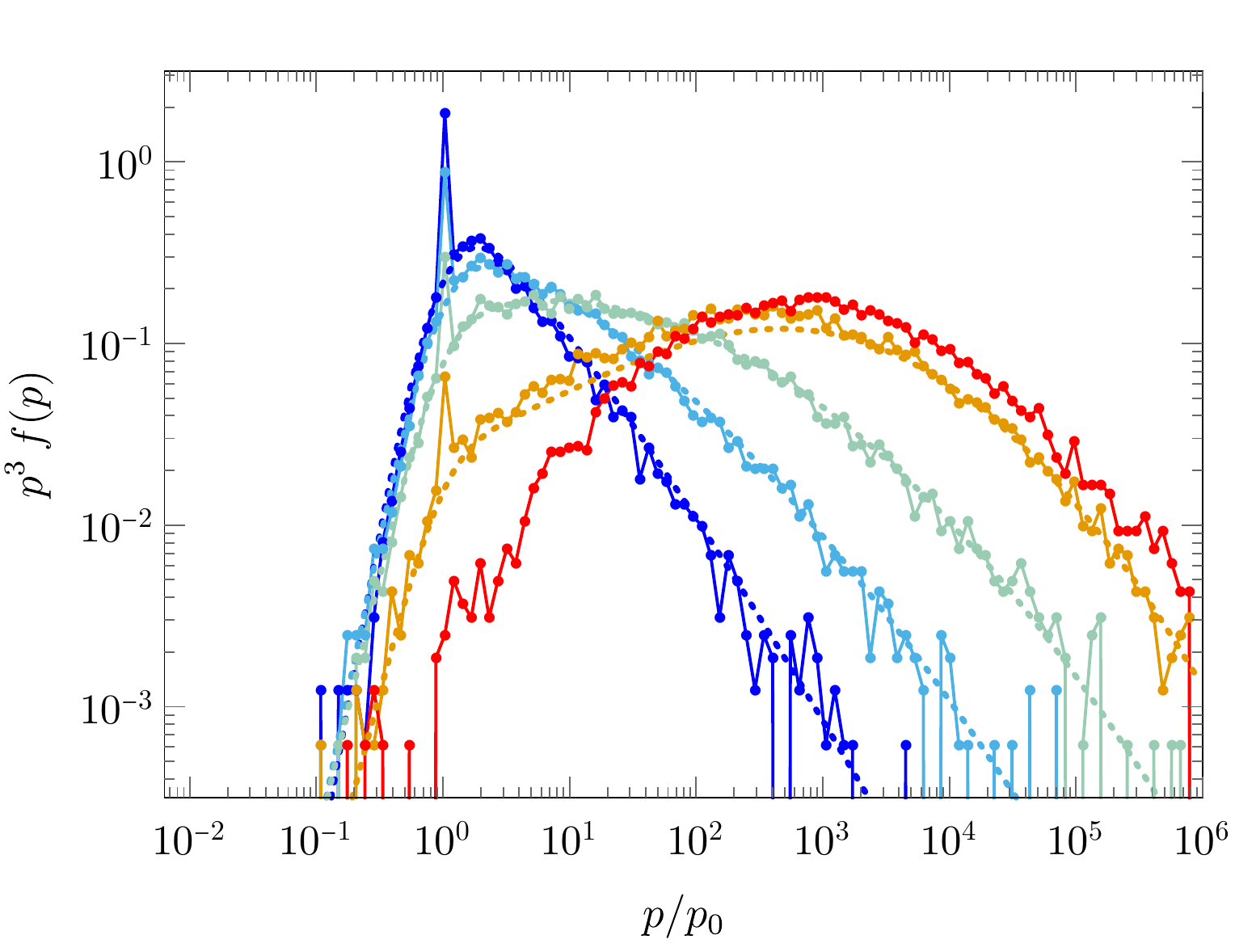}
 \caption{Distributions $p^3f(p,\,t)$ as a function of $p/p_0$ for random walks with L\'evy distributed waiting times, as characterized by the parameters $\alpha=0.3,\,0.5,\,0.7,\,0.9$ (from soft to hard, or blue to orange) at time $\nu_{\rm acc}t=2.5$, with $\Delta t\nu_{\rm acc}=0.1$. The dotted lines show the analytical approximation Eq.~(\ref{eq:solL}) for $k_{\rm max}=3$. In red, we show the spectrum for a gaussian law of waiting times, which can be expressed as a symmetric L\'evy-stable distribution of parameter $\alpha=2$ (here truncated to positive values of the argument). For this case, we recover the standard pile-up distribution at large momenta.
 \label{fig:num22} }
\end{figure}

\begin{figure}
\includegraphics[width=0.45\textwidth]{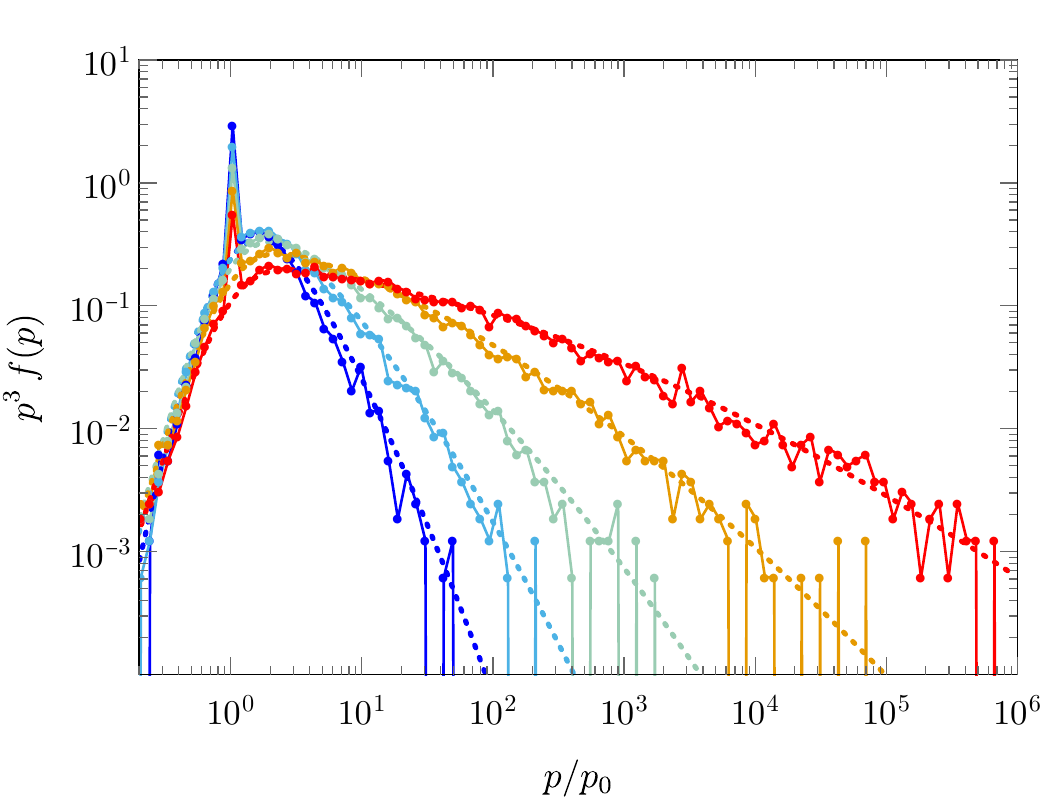}
 \caption{Spectra for the L\'evy random walk model, plotted for a given value $\alpha=0.5$ at different times $\nu_{\rm acc}t\,=\,0.2,\,0.4,\,1.,\,2.5,\,6.3$ (ordered from blue to red), with $\Delta t\nu_{\rm acc}=0.1$. The dotted lines show the analytical approximation Eq.~(\ref{eq:solL}) for $k_{\rm max}=3$. As expected, the spectra become harder as the ratio $t/\Delta t$ increases.
 \label{fig:num23} }
\end{figure}

Formula~(\ref{eq:prob1}) remains valid in the present case, and the probability ${\rm d}P_p(p,n)/{\rm d}p$ is still given by the propagator $G(p\vert p_0,n\Delta t)$. We compute the probability of the particles executing at least $n$ energy gaining jumps within an interval $t$ as the probability of having the sum of the $n$ waiting times less than $t$. The sum of $n$ variables, each distributed according to $L_\alpha(\hat t)$, is distributed as $n^{-1/\alpha}L_\alpha\left(\hat t/n^{1/\alpha}\right)$. Consequently, the probability of achieving at least $n$ jumps within $t$ can be written as:
\begin{align}
P_n(\geq n,\,t)&\,=\, \int_0^{\hat t}{\rm d}\hat\tau\, \frac{1}{n^{1/\alpha}}L_\alpha\left(\frac{\hat\tau}{n^{1/\alpha}}\right)\nonumber\\
&\,=\, \int_0^{\hat t/n^{1/\alpha}}{\rm d}\hat\tau L_\alpha(\hat\tau)\,.
\label{eq:Pn4}
\end{align}
In the large-$n$ limit, we obtain the probability $P_n(n,\,t)$ from (minus) the derivative with respect to $n$, which gives
\begin{equation}
P_n(n,\,t)\,\simeq\, \frac{1}{\alpha}\,\hat t\, n^{-1-\frac{1}{\alpha}}L_\alpha\left(\hat t\, n^{-\frac{1}{\alpha}}\right)\,.
\end{equation}
Approximating the discrete sum in Eq.~(\ref{eq:prob1}) with an integral, we obtain
\begin{align}
p^2 f(p,\,t)\,\simeq\,& \int_{1}^{+\infty}{\rm d}n\,\frac{1}{\alpha}\,\hat t\,n^{-1-\frac{1}{\alpha}}\,L_\alpha\left(\hat t\,n^{-\frac{1}{\alpha}}\right)\,G_{\nu_{\rm acc}}\left(p\vert p_0,n\Delta t\right)\,.
\label{eq:pfL1}
\end{align}
We can obtain useful approximations to this expression by changing variables for $u\,=\,n^{-1/\alpha}\hat t$, and breaking the integrals into two parts, one over the integral $u\in[0,1]$, the other over $u\in[1,\hat t]$ and using both the small and large argument approximations for $L_\alpha(x)$~\citep{Mikusinski59,2010PhRvL.105u0604P,2011PhRvE..84b6702S}:
\begin{align}
L_\alpha(x)&\underset{x\ll1}{\approx}\,\frac{\alpha^{\frac{1}{2(1-\alpha)}}}{\sqrt{2\pi(1-\alpha)}}x^{-\frac{2-\alpha}{2(1-\alpha)}}\exp\left[-(1-\alpha)\alpha^\frac{\alpha}{1-\alpha}\,x^{-\frac{\alpha}{1-\alpha}}\right]\,\nonumber\\
L_\alpha(x)&\underset{x\gg1}{\approx}\,\frac{1}{\pi}\sum_{k=1}^{k_{\rm max}}\,\frac{(-1)^{k+1}}{k!}x^{-1-\alpha k}\Gamma\left(1+\alpha k\right)\sin\left(\pi\alpha k\right)\,.
\label{eq:Lapprox}
\end{align}
With $k_{\rm max}\rightarrow +\infty$, the sum converges to $L_\alpha(x)$ for all $x$, but convergence is slow at $x\,\ll\,1$ and the former expression is more useful. For the calculations that follow, it suffices to choose $k_{\rm max}\sim 3$ in practice. Define the integral
\begin{equation}
I_{u_0,u_1}\left(\kappa,\,\mu,\,\nu,\,\rho\right)\,\equiv\,
\int_{u_0}^{u_1}{\rm d}u\,u^\kappa\,\exp\left[-\mu u^{-\alpha}-\nu u^{\alpha} -
\rho u^{-\frac{\alpha}{1-\alpha}}\right]\,,
\label{eq:defInt1}
\end{equation}
as well as $\hat\nu=\nu_{\rm acc}\Delta t$ and $q_p=\log(p/p_0)^2/(4\hat\nu)$. We obtain
\begin{align}
p^2 f(p,\,t)\,\approx\,&\frac{\left(p/p_0\right)^{1/2}}{\sqrt{4\pi\hat\nu}}\hat t^{-\frac{\alpha}{2}}\,\Biggl\{\nonumber\\
& \quad I_{0,1}\left(\frac{\alpha}{2}-\frac{2-\alpha}{2(1-\alpha)},\, \frac{9}{4}\hat\nu\hat t^\alpha,\,q_p\hat t^{-\alpha},\, (1-\alpha)\alpha^\frac{\alpha}{1-\alpha}\right)\nonumber\\
& \quad +
\frac{1}{\pi}\sum_{k=1}^{k_{\rm max}}\,\frac{(-1)^{k+1}}{k!}\Gamma\left(1+\alpha k\right)\sin\left(\pi\alpha k\right)\nonumber\\
&\quad\quad\times\,I_{1,\hat t}\left(\frac{\alpha}{2}-1-\alpha k,\,\frac{9}{4}\hat\nu\hat t^\alpha,\,q_p\hat t^{-\alpha},0\right)\Biggr\}\,.
\label{eq:solL}
\end{align}
Although the expression appears cumbersome, it boils down to the evaluation of a few integrals and offers a convenient expression for the resulting spectrum, as shown in Figs.~\ref{fig:num22} and \ref{fig:num23}. Figure~\ref{fig:num22} shows some spectra of accelerated particles for $\alpha\in\left\{0.3,\,0.5,\,0.7,\,0.9,\,2.\right\}$ at a given time $\nu_{\rm acc}t=2.5$. The overall shape is close to a powerlaw for small values of $\alpha$, as expected at early times. In this figure, we also show the distribution obtained for a Gaussian distribution of waiting times, which corresponds to the symmetric stable distribution with parameter $\alpha=2$ (here truncated to positive values of the argument). For this case, the solution converges to the standard propagator of the Fokker-Planck equation without traps. As $t$ increases, the functional shape of the distribution function evolves in a non-trivial way for all values of $\alpha$, as illustrated in Fig.~\ref{fig:num23}.

We note that the integral over the interval $[0,1]$ provides the scaling at large $p$, while that over $[1,\hat t]$ determines the low $p$ behaviour. In principle, one can further approximate these integrals, {\it e.g.}, through steepest descent, but no strict powerlaw emerges from the resulting expression. The general scaling is that of an exponential of some power of $\ln(p/p_0)$, which reproduces the rough powerlaw scaling seen in Figs.~\ref{fig:num22} and \ref{fig:num23}.

\section{Discussion}\label{sec:disc}
\subsection{General remarks}
\subsubsection{Particle injection}
Our discussion has tacitly assumed a monoenergetic injection of the form $\delta(p-p_0)$. Our results can nonetheless be extrapolated to other forms of injection by  convolution over the injection kernel. Given that the convolution of a powerlaw with a gaussian does not modify the powerlaw shape, our conclusions remain essentially unaffected for Maxwellian-type injection distributions.

An interesting question that arises is whether the powerlaw distributions that appear in PIC simulations could be due to the injection process itself, rather than to  segregation in stochastic acceleration, as we have proposed. In this picture, one needs to distinguish the thermal pool from the suprathermal particle population which undergoes stochastic acceleration. If particles from the former population are promoted to the latter with a varying degree of success, one could, in principle, observe non-standard -- that is, with respect to standard stochastic acceleration --  distributions at late times. In this regard, we note that convoluting the Green function $G_\nu(p\vert p_0,\,\Delta t)$ of Eq.~(\ref{eq:Green1}) for energy gain through stochastic acceleration with a powerlaw injection kernel $\propto p_0^{-s}$, leads to a distribution function $p^2f(p,\,t)\propto p^{-s}$.

As attractive as it may appear, this interpretation does not seem to fit the numerical experiments reported in \citet{17Zhdankin,18Zhdankin,2018MNRAS.474.2514Z,2018ApJ...867L..18Z}, \citet{18Comisso,2019ApJ...886..122C} and \citet{2020ApJ...893L...7W}. These simulations indeed show that the bulk of the plasma, rather than a fraction of it, is heated by the turbulent cascade and that the mean energy of the bulk population increases with time. The distribution of suprathermal particles appears as a powerlaw extension out of that thermal distribution. Furthermore, the analysis of particle histories reveals that those two states of energization can be attributed to two different processes. Most particles are pre-energized to the thermal bulk by the action of an electric field that is parallel to the mean magnetic field, while most particles in the suprathermal tail are energized by the action of a perpendicular electric field. The former is interpreted as reconnection in small-scale current sheets, while the latter is characteristic of Fermi-type (here, stochastic) acceleration. Appropriately, the former stage can be described as a regular acceleration process, while in the latter stage, particles appear to diffuse in momentum space, with a diffusion coefficient $D_{pp}\propto p^2$, as already mentioned.

\subsubsection{Other approaches}
In principle, the Fokker-Planck equation can contain both diffusion and advection terms, {\it e.g.}
\begin{equation}
    \frac{\partial}{\partial t}f(p,\,t)\,=\,-\frac{1}{p^2}\frac{\partial}{\partial p}
    \left\{A_p p f(p,\,t)\right\}
    + \frac{1}{p^2}\frac{\partial}{\partial p}
    \left\{D_{pp} p^2 \frac{\partial}{\partial p}f(p,\,t)
    \right\}\,,
\label{eq:gFPeq}
\end{equation}
with $A_p$ the advection coefficient, here carrying the same dimensions as $D_{pp}$. $A_p$ may of course depend on momentum; for simplicity, we assume here $A_p\propto p^2$, in accordance with $D_{pp}\propto p^2$ as before and write: $A_p=\nu_{\rm adv}p^2$. This Fokker-Planck equation has a solution, see \cite{2006ApJ...647..539B}, which generally retains a pile-up form. In particular, the mean momentum evolves as 
\begin{equation}
\langle p\rangle\,=\,p_0\,\exp\left[\left(\nu_{\rm adv}+4\nu_{\rm acc}\right)t\right]\,.
\label{eq:pmeangFP}
\end{equation}
Its evolution is thus exponential in time for our choice unless the advection is strongly negative, $A_p=-4D_{pp}$. For such a choice of transport coefficients, one can show that the stationary distribution function scales according to $p^2f_{\rm s}(p)\propto p^{-1}$ at $p\ll p_0$ and $p^2f_{\rm s}(p)\propto p^{-2}$ at $p\gg p_0$. Therefore, a powerlaw shape is preserved, but most particles accumulate at low momenta.

In order to interpret the results of their PIC simulations of stochastic acceleration, \cite{2020ApJ...893L...7W} have measured the advection and diffusion coefficients and shown that the numerical solution of the Fokker-Planck equation determined with those coefficients reproduce satisfactorily the observed spectra. In their case, $D_{pp}\propto p^2$ but the advection coefficient has a non-trivial sign (positive at low momenta, negative at large momenta), a non-trivial energy dependence and its physical origin is not obvious. 

In our model of Sec.~\ref{sec:mod1}, this advection coefficient was set to zero, but the presence of trapping allowed to recover the general powerlaw shape seen in similar numerical simulations. In our view, the present description is more physically motivated than an ad-hoc choice of an advection coefficient, and it also offers a simpler way of extracting physical solutions. For reference, we remark that $A_p=0$ matches the prediction of quasilinear theory in the diffusion approximation \citep{1989ApJ...336..243S}.

Regarding the L\'evy random walk model, most studies consider L\'evy jumps for the momentum (or for the position, when spatial transport is considered) with fixed jumps in time, see for instance \cite{2013ApJ...778...35Z}, \cite{2017ApJ...849...35I} and \cite{2017PhRvL.119d5101I}. In Sec.~\ref{sec:mod2}, we have rather considered momentum jumps characterized by a diffusive propagator, with waiting times distributed according to a L\'evy distribution. Both choices are possible, in principle, but their physical meanings differ. In stochastic acceleration, the typical momentum gain is $\left\langle\Delta p\right\rangle \sim u^2 p$ per scattering event, $u$ denoting the velocity of the scattering center. Hence, a L\'evy walk in momentum space at fixed time intervals might represent a situation in which the velocities are distributed according to some heavy-tailed distribution.  

Such models typically produce hard powerlaws. To see this, consider Eq.~(\ref{eq:prob1}) with jumps of fixed size in time, for instance $P_n(n;t)=\delta\left(n - \nu_{\rm acc}t\right)$. We assume that at each jump, the log-momentum changes by an amount $\Delta\ln p = \Delta\ln p_0\, \hat\Delta$, where $\hat\Delta$ is distributed as $L_\beta(\hat\Delta)$ and $\Delta\ln p_0$ is a reference scale for jumps in momentum. Consequently,  ${\rm d}P_p(p,\,n)/{\rm d}p = p^{-1}\Delta\ln p_0\, \beta^{-1} n^{-1/\beta}\hat\Delta\,L_\beta \left(n^{-1/\beta}\hat\Delta\right)$, which is strongly suppressed at momenta such that $\hat\Delta \lesssim n^{1/\beta}$ and which scales as $p^{-1} \ln (p/p_0)^{-1/\alpha}$ at larger momenta. Such distributions cannot therefore reproduce the soft powerlaws seen in numerical simulations. 

\subsection{Consequences for phenomenology}\label{sec:cons}
So far, our discussion has concerned the time evolution of the distribution function for particles subject to stochastic acceleration only, without considering the possible impact of escape, or even energy losses. Such loss terms nevertheless play an important in shaping the spectra in phenomenological applications, see {\it e.g.}, ~\citet{1984A&A...136..227S} or ~\citet{2008ApJ...681.1725S}. Without entering into the details, we wish to discuss here how the above distribution functions evolve on long time scales, once possible escape losses are considered. We assume, for simplicity, that the scattering timescale of the particles is independent of momentum, a choice consistent with our scaling $D_{pp}\propto p^2$. This implies that escape can be characterized by a momentum-independent scattering frequency $\nu_{\rm esc}$.

\subsubsection{Stochastic acceleration in the presence of traps and escape}
The model that we have developed in Sec.~\ref{sec:mod1} can be directly generalized to this case. Consider for instance the analytical solution Eq.~(\ref{eq:ansolf}). To account for escape losses, we include a distribution $f_{\rm esc}$ that characterizes the population of particles that have escaped the system and rewrite Eq.~(\ref{eq:sysan}) as follows,
\begin{align}
\frac{\partial}{\partial t} f_{\rm esc}(p,\,t)&\,=\,+\nu_{\rm esc,0}f_0(p,\,t) + \nu_{\rm esc,1}f_1(p,\,t),\nonumber\\
\frac{\partial}{\partial t} f_0(p,\,t)&\,=\,-\left(\nu_{01}+\nu_{\rm esc,0}\right)f_0(p,\,t) + \nu_{10}f_1(p,\,t),\nonumber\\
\frac{\partial}{\partial t} f_1(p,\,t)&\,=\,+\nu_{01}f_0(p,\,t) - \left(\nu_{10}+\nu_{\rm esc,1}\right)f_1(p,\,t) \nonumber\\
&\quad + 
\frac{1}{p^2}\frac{\partial}{\partial p}
\left\{p^2D_{pp}\, \frac{\partial}{\partial p} f_1(p,\,t)\right\}\,.\nonumber\\
&
\label{eq:sysan2}
\end{align}
Then, redefining $f_{0/1}(p,\,t) = \exp\left(-\nu_{\rm esc,0/1}t\right)\,g_{0/1}(p,\,t)$, the functions $g_0(p,\,t)$ and $g_1(p,\,t)$ obey the original system (\ref{eq:sysan}), so that their solution is given by Eq.~(\ref{eq:ansolf}), while
\begin{equation}
f_{\rm esc}(p,\,t)\,=\,\int_0^t{\rm d}\tau\,\left[\nu_{\rm esc,0}e^{-\nu_{\rm esc,0}\tau}g_0(p,\tau)+
\nu_{\rm esc,1}e^{-\nu_{\rm esc,1}\tau}g_1(p,\tau)\right]\,,
\label{eq:solfesc}
\end{equation}
and the integral can be carried out explicitly in Eq.~(\ref{eq:ansolf}) to obtain an  expression for $f_{\rm esc}$ that is similar to that for $f_0$ and $f_1$. In the following, we assume $\nu_{\rm esc,0}=\nu_{\rm esc,1}=\nu_{\rm esc}$ for simplicity.

We can expect the following behaviour. Consider first the limit $\nu_{\rm esc}\,\gg\,\nu_{01}$. Then $f_0(p,\,t) \ll f_1(p,\,t)$ and $f_0(p,\,t)\ll f_{\rm esc}(p,\,t)$, because particles have a larger probability of escaping the system than being trapped for some time in the $f_0$ population. This situation is thus similar to that discussed in Sec.~\ref{sec:mod1} with $\nu_{01}^{(0)}\rightarrow 0$, since particles that escape the system never reenter it. We introduce the superscript $^{(0)}$ to index parameters and distribution functions in the absence of escape losses (Sec.~\ref{sec:mod1}). We can obtain an approximate solution from Eq.~(\ref{eq:f0f1s}), provided we make the substitutions $\nu_{10}^{(0)}\rightarrow\nu_{\rm esc}$ and $f_0^{(0)}\rightarrow f_{\rm esc}$. Its powerlaw index becomes a function of the ratio $\nu_{\rm esc}/\nu_{\rm acc}$ and the powerlaw shape holds at all times, even when $\nu_{\rm esc}t\,\gg\,1$.

Consider now the limit $\nu_{\rm esc}\ll\nu_{01}$. We assume $\nu_{01}<\nu_{10}$ and $\nu_{01}<\nu_{\rm acc}$ as before. In this case, we need to consider three regimes, $t\ll\nu_{01}^{-1}$, $\nu_{01}^{-1}\ll t\ll\nu_{\rm esc}^{-1}$ and $t\gg \nu_{\rm esc}^{-1}$. In the first two, $f_{\rm esc}\ll f_0+f_1$, since $\nu_{\rm esc}t\ll1$. The behaviour of $f_0$ and $f_1$ is adequately described by our earlier solution, Eq.~(\ref{eq:ansolf}), given that escape losses can be neglected on those early timescales. Correspondingly, the solution is a powerlaw at early times ($\nu_{01}t\ll1$) and it evolves toward an evolving pile up distribution at late times ($\nu_{01}t\gg1$). 

At late times, $f_1\,\ll\,f_0$ and $f_0$ becomes a pile up distribution, as we have seen (Fig.~\ref{fig:ana2}). Therefore, the present system including $f_{\rm esc}$ is similar to that studied previously, if we make the substitutions $\nu_{01}^{(0)}\rightarrow 0$, $\nu_{10}^{(0)}\rightarrow \nu_{\rm esc}$, $\nu_{\rm acc}^{(0)}\rightarrow \nu_{\rm acc}/\left(1+\nu_{10}/\nu_{01}\right)$, $f_0^{(0)}\rightarrow f_{\rm esc}$ and $f_1^{(0)}\rightarrow f_0$. The replacement $\nu_{\rm acc}^{(0)}\rightarrow  \nu_{\rm acc}/\left(1+\nu_{10}/\nu_{01}\right)$ must be introduced because, as $\nu_{01}t\gg 1$, acceleration proceeds with a rate that is effectively $\nu_{\rm acc}/\left(1+\nu_{10}/\nu_{01}\right)$, not $\nu_{\rm acc}$, see Eq.~(\ref{eq:dNdp2}). Consequently, for $\nu_{\rm esc}t\ll1\ll\nu_{01}t$, we expect to recover a pile up distribution for $f_0$ and a powerlaw for $f_1$, in accord with Eq.~(\ref{eq:f0f1s}). With the above substitutions, this implies a powerlaw for $p^2f_{\rm esc}(p,\,t)$ of exponent $1/2-\left[9/4+\nu_{\rm esc}/\nu_{\rm acc}\left(1+\nu_{10}/\nu_{01}\right)\right]^{1/2}$.  At later times, $\nu_{\rm esc}t\gg1$, $f_0$ disappears and only the powerlaw for $f_{\rm esc}$ remains.

\begin{figure}
\includegraphics[width=0.45\textwidth]{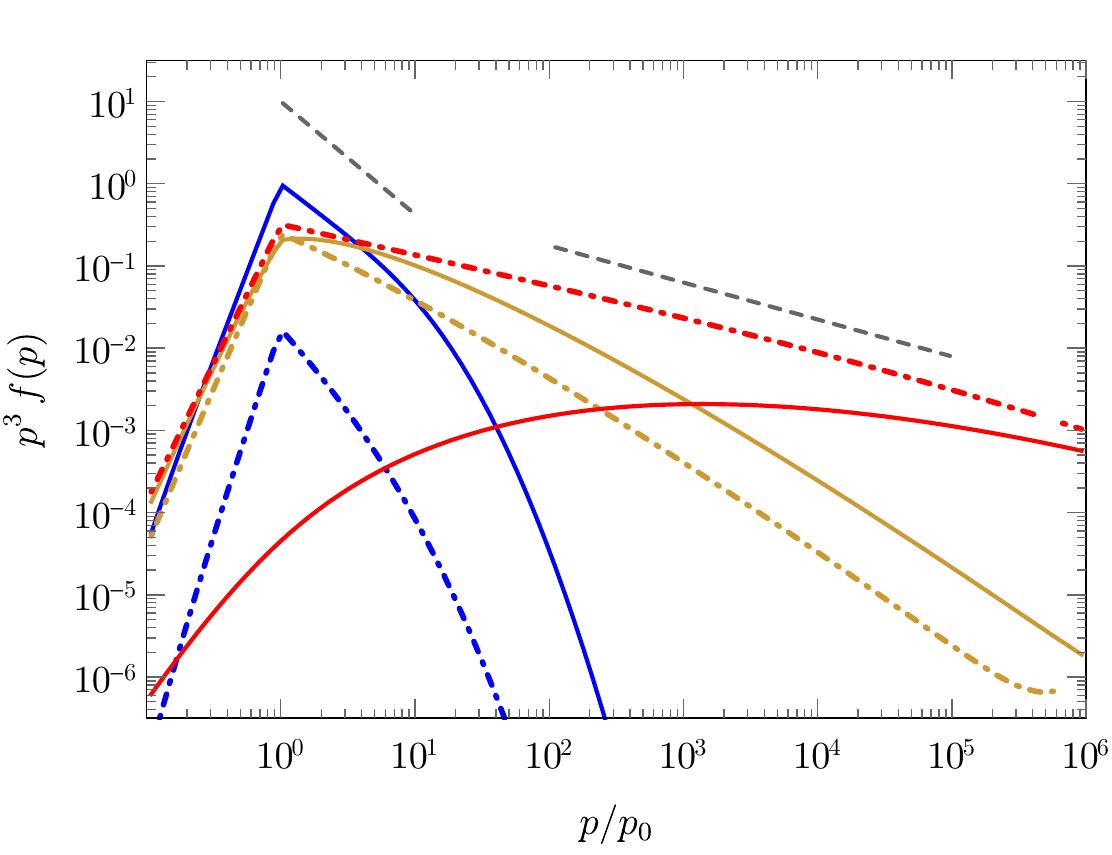}
 \caption{Distributions $p^3f(p,\,t)$, for $f(p,\,t)=f_0(p,\,t)+f_1(p,\,t)$ (solid lines) and $f(p,\,t)=f_{\rm esc}(p,\,t)$ (dash-dotted lines) at different times, ordered from blue to red (or left to right): $\nu_{\rm acc}t=0.4,\,10.,\,80.$, for $\nu_{01}/\nu_{\rm acc}=0.2$, $\nu_{10}/\nu_{\rm acc}=6.$ and $\nu_{\rm esc}/\nu_{\rm acc}=0.05$. The distribution function of escaped particles $f_{\rm esc}(p,\,t)$ scales at all times as a powerlaw. The dashed gray lines indicate the powerlaw escaped at early time, as predicted by Eq.~(\ref{eq:f0f1s}), and at late times, as discussed in the text. 
 \label{fig:esc1} }
\end{figure}

We illustrate this behaviour with Fig.~\ref{fig:esc1}, which shows the time evolution of $f_0+f_1$ and $f_{\rm esc}$ for the choice $\nu_{\rm acc}t=0.4,\,10.,\,80.$, for $\nu_{01}/\nu_{\rm acc}=0.2$, $\nu_{10}/\nu_{\rm acc}=6.$ and $\nu_{\rm esc}/\nu_{\rm acc}=0.05$. At early times, both $f_{\rm esc}$ and $f_0+f_1$ scale as powerlaws, whose index is given by Eq.~(\ref{eq:f0f1s}), as expected. At late times, $f_0+f_1$ evolve toward pile up distribution, while $f_{\rm esc}$ retains a powerlaw shape, but with a different index, close to that predicted above in terms of $\nu_{\rm esc}/\nu_{\rm acc}$ and $\nu_{10}/\nu_{01}$.

\subsubsection{L\'evy random walks including escape terms}
Consider now a L\'evy random walker, including the possibility of escape at frequency $\nu_{\rm esc}$. This means that, at each jump in momentum, $p\rightarrow p+\Delta p$, the particle has a probability $p_{\rm esc}=1-\exp(\nu_{\rm esc}\hat t\Delta t)$ of escaping the system, where $\hat t$ is distributed according to the stable distribution $L_\alpha(\hat t)$. This effect can be easily included in the discretized random walk, and its impact on the spectrum can be estimated as follows.

At early times, $\nu_{\rm esc}t\,\ll\,1$, escape plays little role and it can be neglected. The shape of the spectrum is therefore not modified with respect to that obtained in Eq.~(\ref{eq:solL}). At late times, $\nu_{\rm esc}t\,\gg\,1$, escape is bound to shape the spectrum and to turn it into an approximate powerlaw. A first approximation for its exponent can be obtained from Eq.~(\ref{eq:pstat}), namely $-1+\ln(1-p_{\rm esc})/g$, where $g\simeq 4\nu_{\rm acc}\Delta t$ represents $\Delta\ln p$. The escape probability is given by
\begin{equation}
p_{\rm esc}\,\simeq\,\int_{\left(\nu_{\rm esc}\Delta t\right)^{-1}}^{+\infty}{\rm d}\hat t\,L_\alpha(\hat t)\,\simeq\, \frac{1}{\pi}\left(\nu_{\rm esc}\Delta t\right)^\alpha\Gamma(\alpha)\sin(\alpha\pi)\,.
\label{eq:pescL}
\end{equation}
It thus depends on $\alpha$, giving a spectrum that is harder with increasing $\alpha$.

\begin{figure}
\includegraphics[width=0.45\textwidth]{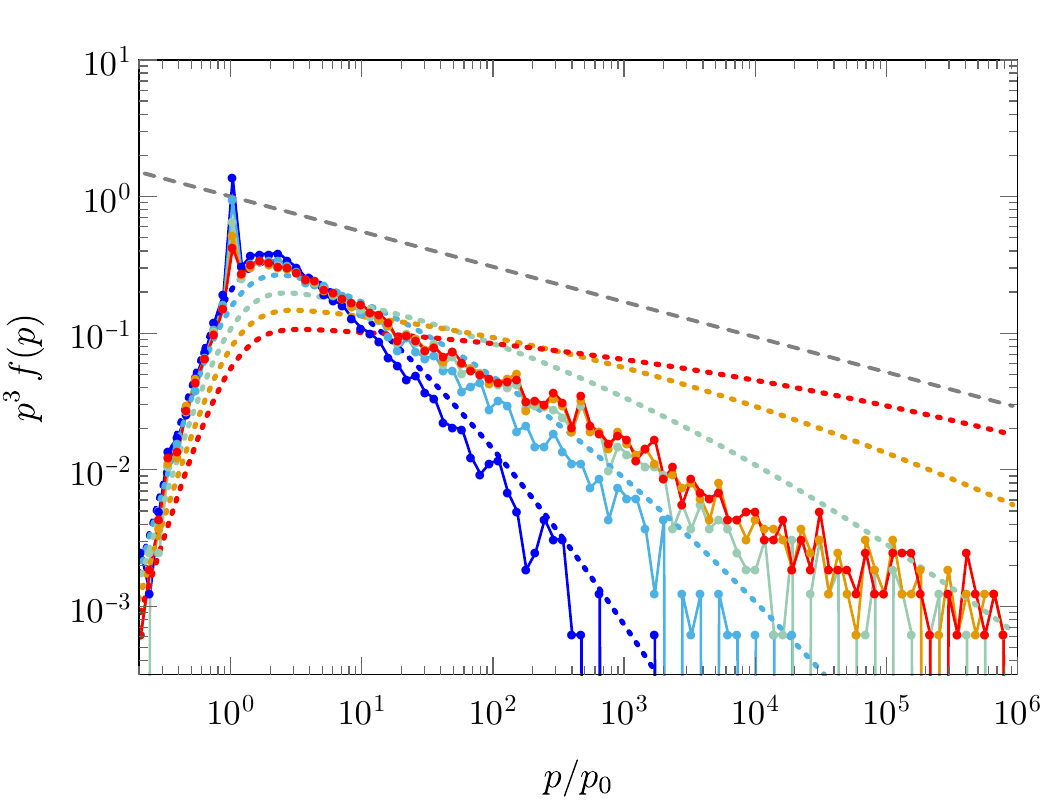}
 \caption{Evolution in time (ordered from blue to red, or left to right) of the distribution function for a random walk in momentum space with L\'evy waiting times, for $\alpha=0.5$, $\nu_{\rm acc}\Delta t=0.1$ and $\nu_{\rm acc}t\,=\,1.0,\,2.5,\,6.3,\,13,\,28$, including escape at frequency $\nu_{\rm esc}=0.3\nu_{\rm acc}$. The dotted lines represent the analytical solutions without escape, while the dashed line shows the powerlaw with the index discussed in the text, function of $\alpha$ and $\nu_{\rm esc}$. At late times, the numerical solution converges to a powerlaw, with an index slightly steeper (exponent $\simeq-1.5$ for $p^2f$) than our simple estimate ($-1.3$).
 \label{fig:escL} }
\end{figure}

In Fig.~\ref{fig:escL}, we plot the evolution of the distribution function in the presence of escape, for the case $\alpha=0.5$. As before, we assume $\nu_{\rm acc}\Delta t=0.1$, and we choose here $\nu_{\rm esc}=0.3\nu_{\rm acc}$. At early times $\nu_{\rm acc}t=\,1,\,2.5$ corresponding to $\nu_{\rm esc}t<1$, the solution agrees with the analytical solution without escape, as anticipated. At later times, $\nu_{\rm esc}t>1$, while the distribution for $\nu_{\rm esc}=0$ [Eq.~(\ref{eq:solL})] departs from a powerlaw form and becomes harder and harder, the numerical solution that considers finite escape losses converges to a powerlaw with an index that is not very different from our prediction ($-1.5$ measured for $\alpha=0.5$ {\it vs} $-1.3$ predicted by the above estimate).

\section{Conclusions}\label{sec:conc}
This paper has discussed the physics of stochastic particle acceleration using continuous-time random walks, in which the time span that separates two energy-jump events is distributed as a continuous random variable. This study is motivated by the result of recent numerical simulations of particle acceleration in magnetized turbulence, which have produced powerlaw spectra where pile-up distributions were theoretically expected. As we have argued in Sec.~\ref{sec:introd}, such an observation is an indication for the existence of some ``trapping'', which inhibits acceleration for some of the particles, and as such, acts as a form of escape on the finite timescale of those simulations. The powerlaw then results from the competition between energy gain and escape/trapping, a common trait of Fermi-type acceleration.

This segregation of particles is likely related to a non-trivial dependence on the acceleration rate on phase space variables other than the momentum, {\it e.g.} the pitch-angle of the particle, or its spatial position. Both dependencies are indeed averaged out when one considers momentum diffusion only. Our description of stochastic acceleration in terms of continuous-time random walks provides a simple way to describe the consequences of such hidden dependencies.

In Sec.~\ref{sec:mod1}, we have discussed random walks with finite mean waiting time, considering in particular distributions characterized by two timescales of acceleration, one slow and one fast. We have shown that a powerlaw indeed emerges as a natural consequence of stochastic acceleration if the timescale on which one probes the distribution function, {\it e.g.} the simulation timescale, is shorter than the slow timescale. The slope of the powerlaw can then be expressed in terms of the fast acceleration timescale and of the typical time over which a given particle transits into the region of phase space where acceleration takes place on the slow timescale.  On longer timescales, the distribution of accelerated particles converges to a pile-up distribution, as expected, albeit with an effective acceleration timescale which is significantly enlarged by the amount of time spent in traps. We have provided a general analytical solution for the distribution function as well as simplified analytical estimates in both limits.

In Sec.~\ref{sec:mod2}, we have discussed the other general class of continuous-time random walks, that of heavy-tailed distributions of waiting time, with infinite mean. We have considered in particular one-sided L\'evy-stable distributions, which behave as attractors for that class of distribution functions. Here as well, we have provided analytical estimates which match dedicated numerical Monte Carlo simulations of the stochastic process. By construction, one cannot define here a slow and a fast timescale. The distribution cannot therefore be fully described by a powerlaw at high energies, although the running of the powerlaw exponent with momentum is rather mild. As one waits longer and longer, the distribution becomes harder and harder, until the mean momentum itself starts to increase, the distribution then turning into a pile-up form.

Our study thus provides a simple interpretation of the observation of powerlaws in recent numerical simulations and it clearly highlights the need for an improved understanding of the possible hidden dependencies of the acceleration rate. If confirmed by future numerical experiments, the shape and time dependence of the accelerated distribution could be used to characterize the distribution of waiting times. Our results can be generalized and applied to concrete astrophysical scenarios, by adding in the possible influence of energy losses, escape losses etc. As an illustration, we have discussed the influence of escape losses assuming a momentum-independent scattering timescale, and shown that such losses lead to a softened powerlaw distribution.

\section*{Acknowledgements}
We thank Luca Comisso, Camilia Demidem and Lorenzo Sironi for insightful discussions.
This work was initiated at the Kavli Institute for Theoretical Astrophysics, University of California, Santa Barbara; it has been supported in part by the National Science Foundation under Grant No.~NSF PHY-1748958. ML acknowledges support by the Sorbonne Universit\'e DIWINE Emergence-2019 program. MM acknowledges support from NASA ATP-program within grant 80NSSC17K0255.

\section*{Data availability}
The data underlying this article are available in the article.

\bibliographystyle{mnras}
\bibliography{turb} 

\appendix

\section{Analytical solution for model 1}\label{sec:appA}
To solve the system of equations~(\ref{eq:sysan}), we perform a Laplace transform in time to write ($\lambda$ denotes the Laplace conjugate variable for $t$)
\begin{equation}
\begin{cases}
\left(\lambda +\nu_{01}\right)\tilde f_0-\nu_{10}\tilde f_1&\,=\, f_{0}^{0}\nonumber\\
\left(\lambda +\nu_{10}\right)\tilde f_1-\nu_{01}\tilde f_0&\,=\, \displaystyle{\frac{1}{p^2}\frac{\partial}{\partial p}\left\{ \nu_{{\rm acc}}p^{4}\frac{\partial}{\partial p}\tilde f_1\right\} +f_1^0}\,,
\end{cases}
\label{eq:sysanL}
\end{equation}
where $f_0^0 = f_0(p,\,t=0)$ and $f_1^0=f_1(p,\,t=0)$. It proves convenient to switch variables from $p$ to $q=p^{-3}$, which leads to
\begin{equation}
\begin{cases}
\displaystyle{\tilde f_0}&\,=\,\displaystyle{\frac{\nu_{10}}{\lambda +\nu_{01}}\tilde f_1+ \frac{1}{\lambda +\nu_{01}}f_{0}^{0}}\nonumber\\
\displaystyle{q^{2}\frac{\partial^2}{\partial q^2}\tilde f_1-\frac{\lambda }{9\nu_{\rm acc}}\left(1+\frac{\nu_{10}}{\lambda +\nu_{01}}\right)\tilde f_1} & =\displaystyle{-\frac{1}{9\nu_{\rm acc}}
\frac{\nu_{01}}{\lambda +\nu_{01}}f_0^0-\frac{1}{9\nu_{\rm acc}}f_1^0}\,,
\end{cases}
\label{eq:sysanL2}
\end{equation}
The Green's function $F(q;\,q')$ such that
\begin{equation}
q^{2}\frac{\partial^2}{\partial q^2}F-\Upsilon\,F\,=\,\delta\left(q-q'\right)\, ,
\label{eq:appLg1}
\end{equation}
where
\begin{align}
\Upsilon(\lambda ) &\,=\,  \frac{\lambda }{9\nu_{{\rm acc}}}\left(1+\frac{\nu_{10}}{\lambda +\nu_{01}}\right)\,,
\label{eq:appL1}
\end{align}
can be expressed as
\begin{equation}
F(q;\,q')\,=\,-\frac{\sqrt{q/q'}}{2q'\sqrt{\Upsilon(\lambda )+\frac{1}{4}}}
e^{-\sqrt{\Upsilon(\lambda )+\frac{1}{4}}\left\vert\ln(q/q')\right\vert}\,.
\label{eq:appf1gr}
\end{equation}
Hence, the Laplace transform of $f_1(p,\,t)$ can be written as
\begin{equation}
\tilde f_1(q,\lambda )\,=\,-\frac{1}{9\nu_{\rm acc}}\int{\rm d}q'\,F(q;\,q')\,\Gamma_1(q';\,\lambda )\,,
\label{eq:appf1sol1}
\end{equation}
with
\begin{equation}
\Gamma_1(q',\lambda ) \,=\, \left(\frac{\nu_{01}}{\lambda +\nu_{01}}f_0^0+f_1^0\right)\,.
\label{eq:defg1}
\end{equation}
The initial distributions are evaluated at $p'=q'^{-1/3}$ in the above expression.
In the following, we consider initial data of the form $\delta(p-p_0)$ and thus operate the substitution $f_0^0\rightarrow f_0^0 p_0\delta (p-p_0)$, $f_1^0\rightarrow f_1^0 p_0\delta (p-p_0)$ to obtain
\begin{equation}
\tilde f_1(q,\lambda )\,=\,\frac{\sqrt{q/q_0}e^{-\sqrt{\Upsilon(\lambda )+\frac{1}{4}}\left\vert\ln(q/q_0)\right\vert}}{6\nu_{\rm acc}\sqrt{\Upsilon(\lambda )+\frac{1}{4}}}\,\Gamma_1(q_0;\,\lambda )\,.
\label{eq:appf1sol2}
\end{equation}

To simplify further the notations, we define $x=\left\vert\ln(q/q_0)\right\vert$. The distribution function $f_1$ is then obtained through the inverse Laplace transform
\begin{equation}
f_{1}\left(p,\,t\right)\,=\,\frac{\sqrt{q/q_{0}}}{12i\pi\nu_{\rm acc}}\int_{L}{\rm d}\lambda \,\frac{e^{\lambda  t-
\sqrt{\Upsilon(\lambda )+\frac{1}{4}} x}}{\sqrt{\Upsilon(\lambda )+\frac{1}{4}}}\Gamma_1(q_0,\lambda )\,,
\label{eq:solf11}
\end{equation}
hence the solution for $f_0$:
\begin{equation}
f_0\left(p,\,t\right)\,=\,e^{-\nu_{01}t}f_0^0\,+\,\frac{\sqrt{q/q_{0}}}{12i\pi\nu_{\rm acc}}\int_{L}{\rm d}\lambda \,\frac{e^{\lambda  t-
\sqrt{\Upsilon(\lambda )+\frac{1}{4}} x}}{\sqrt{\Upsilon(\lambda )+\frac{1}{4}}}\Gamma_0(q_0,\lambda )\,,
\label{eq:solf10}
\end{equation}
with 
\begin{equation}
\Gamma_0(q,\lambda ) \,=\,\frac{\nu_{10}}{\lambda +\nu_{01}}\Gamma_1(q,\lambda )\,=\,\frac{\nu_{01}\nu_{10}}{\left(\lambda +\nu_{01} \right)^2}f_0^0 + \frac{\nu_{10}}{\lambda +\nu_{01}}f_1^0\,.
\label{eq:defg0}
\end{equation}
The Bromwich integrals are of the form
\begin{equation}
I\,=\,\frac{1}{2i\pi}\int_{L}{\rm d}\lambda \,\frac{e^{\lambda  t-
\sqrt{\Upsilon(\lambda )+\frac{1}{4}} x}}{\sqrt{\Upsilon(\lambda )+\frac{1}{4}}}\Gamma(\lambda )\,,
\label{eq:appLI}
\end{equation}
and contain branch cuts on the negative real axis where the argument of the square root $\sqrt{\Upsilon(\lambda )+\frac{1}{4}}$ becomes negative. In detail, 
\begin{equation}
\sqrt{\Upsilon(\lambda )+\frac{1}{4}} \,=\,\frac{1}{3\sqrt{\nu_{\rm acc}}}\sqrt{\frac{(\lambda -\lambda _-)(\lambda -\lambda _+)}{\lambda +\nu_{01}}}\,,
\label{eq:ups}
\end{equation}
with
\begin{align}
\lambda _{\pm}&\,=\,\frac{1}{2}\Biggl\{-\left(\nu_{10}+\frac{9}{4}\nu_{\rm acc}+\nu_{01}\right)\nonumber\\
&\quad\quad\quad\pm\left[
\left(\nu_{10}+\frac{9}{4}\nu_{\rm acc}-\nu_{01}\right)^2 + 4 \nu_{01}\nu_{10}\right]^{1/2}\Biggr\}\,.
\label{eq:zpm}
\end{align}

\begin{figure}
\centering \includegraphics[width=0.35\textwidth]{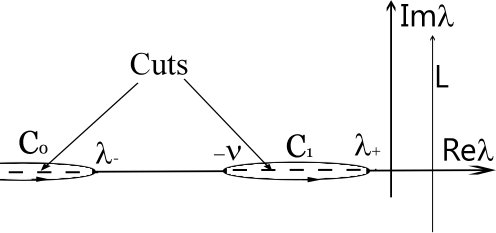}
\caption{Contours around the branch cuts used for the inverse Laplace transform of Eq.~(\ref{eq:appLI}).
\label{fig:Lcontour}}
\end{figure}

Both roots are negative and ordered according to $\lambda _{-}<-\nu_{01}<\lambda _+$. The branch cuts are at $\Re \lambda <\lambda _-$ and $-\nu_{01}<\Re \lambda <\lambda _+$, which gives the two contours of integration $\mathcal C_0$ and $\mathcal C_1$ picture on Fig.~\ref{fig:Lcontour}.
We thus obtain
\begin{equation}
I\,=\,\frac{1}{\pi}\left\{\int_{-\lambda _+}^{\nu_{01}}+\int_{-\lambda _-}^{+\infty}\right\}{\rm d}\lambda \, e^{-\lambda  t}\,\frac{\cos\left[\sqrt{\left\vert\Upsilon(\lambda )+\frac{1}{4}\right\vert}x\right]}{\sqrt{\left\vert\Upsilon(\lambda )+\frac{1}{4}\right\vert}}\Gamma(-\lambda )\,.
\label{eq:appLintI}
\end{equation}
The poles of $\Gamma(-\lambda )$ at $\nu_{01}$ (which appear up to second order in $\tilde f_0$ and $\tilde f_1$) do not provide any additional contribution to the contour.

Finally, changing variables $\lambda  = s \nu_{\rm acc}$, and defining, for the sake of clarity
\begin{equation}
\Sigma(s)\,=\,\sqrt{\frac{\left(s+\frac{\lambda_-}{\nu_{\rm acc}}\right)
\left(s+\frac{\lambda_+}{\nu_{\rm acc}}\right)}{s-\frac{\nu_{01}}{\nu_{\rm acc}}}}
\label{eq:appsigs}
\end{equation}
we obtain:
\begin{align}
f_0\left(p,\,t\right)&\,=\,e^{-\nu_{01}t}f_0^0\nonumber\\
&\quad\,+\,\frac{1}{2\pi}\left(\frac{p}{p_0}\right)^{-3/2}\left\{\int_{-\lambda _+/\nu_{\rm acc}}^{\nu_{01}/\nu_{\rm acc}}+\int_{-\lambda _-/\nu_{\rm acc}}^{+\infty}\right\}{\rm d}s\,
e^{-s \nu_{\rm acc}t}\nonumber\\
&\quad\quad\quad\,\times\,\frac{\cos\left[\Sigma(s)\ln(p/p_0)\right]}
{\Sigma(s)}\,\Gamma_0(-s\nu_{\rm acc})\,\nonumber\\
f_1\left(p,\,t\right)&\,=\,\frac{1}{2\pi}\left(\frac{p}{p_0}\right)^{-3/2}\left\{\int_{-\lambda _+/\nu_{\rm acc}}^{\nu_{01}/\nu_{\rm acc}}+\int_{-\lambda _-/\nu_{\rm acc}}^{+\infty}\right\}{\rm d}s\,
e^{-s \nu_{\rm acc}t}\nonumber\\
&\quad\quad\quad\,\times\,\frac{\cos\left[\Sigma(s)\ln(p/p_0)\right]}
{\Sigma(s)}\,\Gamma_1(-s\nu_{\rm acc})\,.
\label{eq:appLfin}
\end{align}

In the limit $\nu_{01}\,\rightarrow\,0$, $\lambda_+\rightarrow 0$ hence the integral over the contour $\mathcal C_1$ vanishes, and $\lambda_-\rightarrow -\left(\nu_{10} + \frac{9}{4}\nu_{\rm acc}\right)$. Assuming for simplicity $f_0^0=0$, changing variables $y=\Sigma(s)$, we obtain
\begin{align}
f_1\left(p,\,t\right)&\,=\,\frac{1}{\pi}\left(\frac{p}{p_0}\right)^{-3/2}\int_0^{+\infty}{\rm d}y\,e^{-\left(\nu_{10} + \frac{9}{4}\nu_{\rm acc}\right)t-\nu_{\rm acc}t y^2}\nonumber\\
&\quad\quad\quad\times\,\cos\left[y\ln(p/p_0)\right]f_1^0\,\nonumber\\
&\,=\,\frac{1}{2\sqrt{\pi\nu_{\rm acc}t}}\left(\frac{p}{p_0}\right)^{-3/2}\,e^{-\left(\nu_{10} + \frac{9}{4}\nu_{\rm acc}\right)t - \frac{\ln(p/p_0)^2}{4\nu_{\rm acc}t}}f_1^0\,.
\label{eq:f1fin}
\end{align}
The distribution for $f_0$ can be obtained in a similar way, although it proves more convenient to directly integrate Eq.~(\ref{eq:sysan}) in this case. The resulting expression is given in the main text, see Eq.~(\ref{eq:f0f1s}).

\bsp	
\label{lastpage}
\end{document}